# Sub-diffraction Imaging of Carrier Dynamics in Halide Perovskite Semiconductors: Effects of Passivation, Morphology, and Ion Motion

*Madeleine D. Breshears*[a,+], *Justin Pothoof*[a,+], *Rajiv Giridharagopal*[a], *David S. Ginger*[a,*]

[a]Department of Chemistry, University of Washington, Box 351700, Seattle, Washington, 98195-1700, United States

[*]Corresponding Author: dginger@uw.edu

[+]M.D. Breshears and J. Pothoof contributed equally to this work.

**Abstract**

We spatially resolve photocarrier dynamics in halide perovskites using time-resolved electrostatic force microscopy (trEFM) to map surface potential equilibration during photoexcitation. Following treatment with different surface passivation agents, we show that trEFM probes dynamics directly related to surface recombination velocity and carrier lifetimes correlated with time-resolved photoluminescence. Our results reveal nanoscale variations in recombination dynamics following surface passivation. We also observe heterogeneity in surface potential equilibration times dependent on perovskite film morphology. We combine wavelength- and intensity-dependent measurements with drift-diffusion simulations to disentangle the influence of carrier recombination and ion migration on surface potential equilibration. These results demonstrate that we can use mechanical detection to image electronic carrier recombination dynamics in perovskites below the optical diffraction limit while also showing the potential for future improvements in heterogeneity of surface passivation.

## Introduction

Halide perovskite semiconductors have numerous applications including solar cells,[1,2] light-emitting diodes,[2,3] radiation detectors,[4] and even sources of quantum light.[5] Their advantages include scalable additive manufacturing, large absorption coefficients, bandgap tunability,[6] and high defect tolerance.[2] While perovskites are defect tolerant, they are not defect free.[2,7–11] Passivation strategies to reduce surface recombination velocity (SRV) by a factor of ~100 have improved perovskite device efficiencies over the last several years.[7,12–17] At the same time, ion motion, linked to phase segregation and electrochemical reactions, can be a limiting factor in perovskite semiconductor performance and stability.[18–25] Probing these effects at the device level is useful, but inherently reflects average values given the heterogenous nature of halide perovskite thin films,[8,9,26–28] and understanding how local film microstructure correlates with both electronic and ionic defects could advance materials processing.[8,9,27,29–31]

Indeed, microscopy tools that elucidate the relationship between local structure and nonradiative recombination have already proven useful for analyzing and improving semiconductor performance.[8,20,27,29,30,32–35] Time-resolved photoluminescence (trPL) probes electronic carrier recombination processes in perovskite materials;[36,37] and with confocal microscopy, we can obtain spatially resolved trPL maps at the optical diffraction limit. However, current best-in-class perovskites comprise mixed-cation, mixed-halide compositions with grains that are often ~100 nm or smaller,[38–40] which is far below the optical diffraction limit with visible light. Other techniques, such as cathodoluminescence microscopy, can map carrier recombination dynamics on the nanoscale using an electron beam but often damage the sample in the process.[32,41] The ideal microscopy tool for advancing cutting edge perovskites would thus allow for measurement of carrier recombination dynamics at nanometer scale resolution without damaging the sample.

Time-resolved electrostatic force microscopy (trEFM), a scanning probe microscopy technique among the family of methods capable of providing fast electronic dynamics,[42–46] uses a mechanical cantilever to sense dynamic changes in the local electrostatic force gradient between the tip and the sample during photoexcitation.[30,31,47–49] To date, we have used trEFM to measure systems with large exciton binding energies such as organic photovoltaics and layered Ruddlesden-Popper perovskites.[30,31,43,48–53] In mixed-cation, mixed-halide perovskite systems where photoexcitation leads to free carriers, we expect the time-dependent changes measured should readily probe local recombination rates rather than be quantum efficiency-limited.

Here, we compare high-resolution trEFM images with spatially averaged trPL and drift-diffusion simulations. We show that trEFM measures the time it takes for surface potential to equilibrate as controlled by the evolution of electronic and ionic carrier concentrations during photoexcitation, and we probe the effects of chemical passivation on electronic carrier recombination and of local defect-mediated ion migration.

We demonstrate that treatments with different surface passivators, including (3-aminopropyl)trimethoxysilane (APTMS),[14–16] [3-(2-aminoethylamino)propyl]trimethoxysilane (AEAPTMS),[14] and phenethylammonium iodide (PEAI),[54,55] result in slower surface potential equilibration times due to reduced surface recombination velocities. We find that the local carrier dynamics vary with the perovskite morphology, where grain boundaries exhibit slower photoinduced dynamics than grain interiors. Through a combination of wavelength- and intensity-dependent experiments and drift-diffusion simulations, we show the competing influence of electronic carrier recombination and mobile ions on the surface potential equilibration time. We attribute the slower surface potential equilibration times following chemical passivation to the suppression of SRV; we ascribe the slower surface potential equilibration time observed at grain boundaries to locally higher defect densities that mediate slow ion migration and incur increased trap-mediated electronic carrier motion. Importantly, we show that surface potential equilibration times measured by trEFM correlate directly with minority carrier lifetimes and SRVs as measured by trPL. Our results connect the properties measured by trEFM with established optical characterization techniques and verify that trEFM can be a predictive tool for evaluating carrier recombination dynamics in modern halide perovskites with sub-diffraction-limited resolution, while also demonstrating that even common, effective surface passivation schemes can still exhibit local heterogeneity.

## Main

Fig. 1a shows a schematic of the trEFM experiment we use to probe carrier dynamics below the diffraction limit.[48,49] Here, we photoexcite the sample through the ITO substrate using a laser with a rise time of ~2 ns and record the evolution of the mechanical cantilever dynamics to capture the transient electronic response of the sample. Fig. 1b shows the timing diagram of the excitation. Following photoexcitation, the carrier populations generated by the illumination will evolve to a new equilibrium. The time-dependence of these processes leads to a time-dependent change in the surface potential, surface capacitance, and dissipative properties of the sample,[42,43,48,49] and these shifts lead to transient deviations of the cantilever motion from the steady-state sinusoidal motion of a damped, driven harmonic oscillator.[56] These transient deviations encode information about short time dynamics of the local electronic environment's approach to a new equilibrium, which can be reconstructed via empirical calibration,[48] or even simulation and machine learning approaches.[53] Fig. 1c shows a representative cantilever frequency shift over time in response to photoexcitation. We obtain this frequency shift by demodulating the cantilever's oscillation (see Methods and Supplementary Note 1). The rapid shift in the cantilever's frequency following photoexcitation captures the transient dynamics of interest; following that, the cantilever relaxes to its resonance frequency on timescales governed by the cantilever quality factor, Q. We fit the frequency vs. time trace to a product of two exponentials (black dashed line in Fig. 1c), which enables us to find the time it takes for the cantilever to reach its maximum frequency deviation. We calibrate this time against the simulated response to excitation events with a defined time constant to extract a cantilever-independent time constant, $\tau$ (see Methods, Supplementary Note 1, and Supplementary Fig. 1).[48,49] Next, we show that this time constant reflects the surface potential equilibration time in halide perovskites.

The surface potential equilibration time is controlled by the sample's electronic carrier recombination dynamics and illumination conditions. First, we show there is a qualitative correlation between trEFM and confocal PL microscopy over large length scales on a partial perovskite device stack (or "half-stack)" with the architecture ITO/Me-4PACz/$Cs_{0.17}FA_{0.83}Pb(I_{0.75}Br_{0.25})_3$. This perovskite formulation is known to exhibit PL heterogeneity on the multi-micron-scale.[57] Fig. 1d shows a map of surface potential equilibration times in a 17.6×10 μm region-of-interest (ROI) as measured by trEFM, with empirical time constants on the microsecond-scale. Supplementary Fig. 2a shows the topography of this ROI. After imaging the sample with trEFM, we perform correlated confocal PL microscopy (see Methods). Fig. 1e shows the PL intensity map of the ROI. Fig. 1f shows correlated line traces (marked by the dashed lines in Fig. 1d,e), where slower surface potential equilibration times measured by trEFM correspond to brighter PL. Annotations on Fig. 1f show average minority carrier lifetime values collected at the corresponding positions shown in Fig. 1e, confirming that slower surface potential equilibration times measured by trEFM not only correlate to brighter PL, but also, qualitatively longer minority carrier lifetimes. Supplementary Fig. 2b shows a positive correlation between local surface potential equilibration times measured by trEFM and carrier dynamics measured by trPL in this image, with a Pearson correlation coefficient of 0.88 (with a p-value of 0.018).

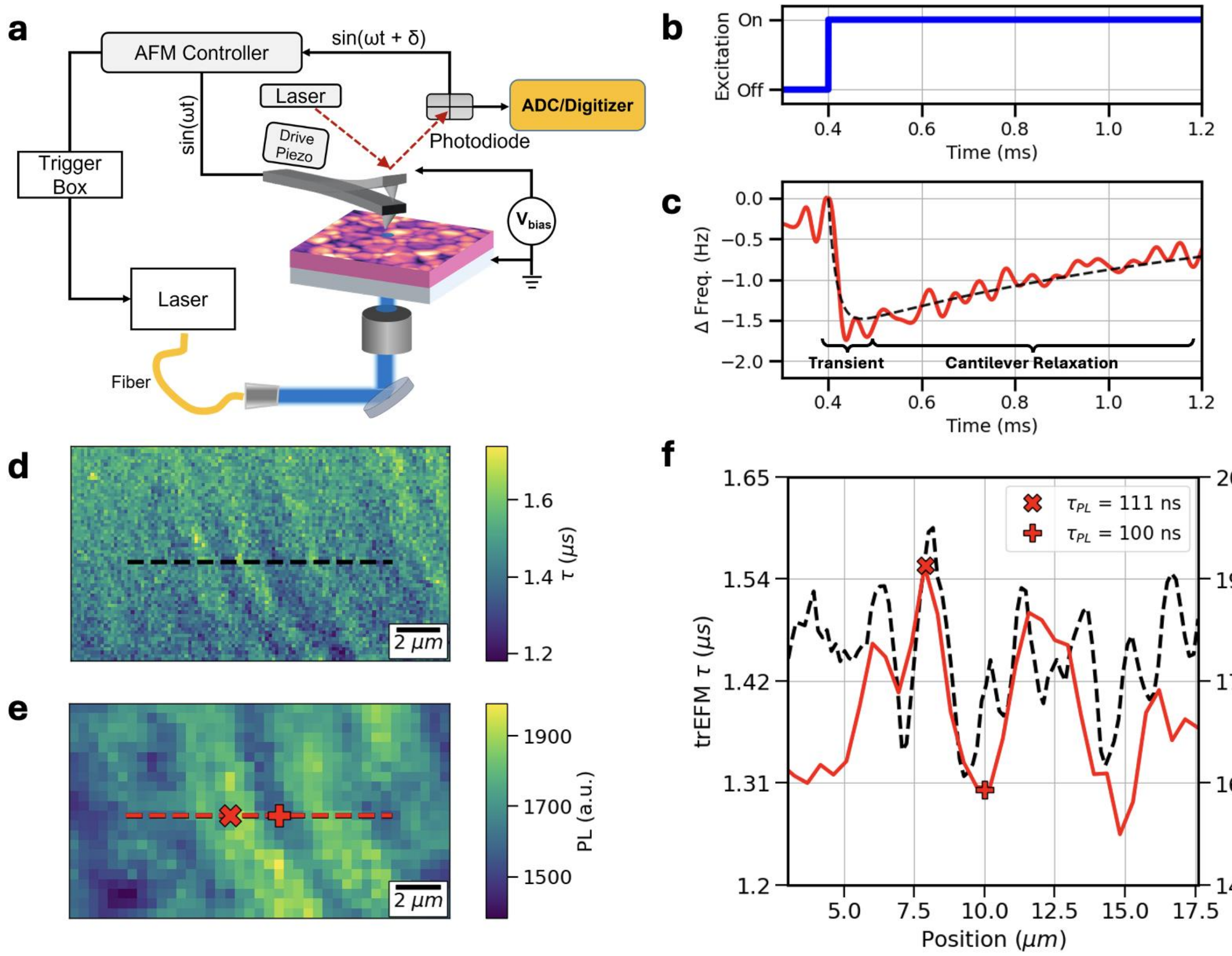

**Fig. 1: trEFM measures photoinduced dynamics that correlate to PL metrics of interest**. (a) trEFM schematic, where ADC refers to an analog-to-digital converter. (b) Excitation protocol for typical trEFM experiment. (c) Change in instantaneous frequency of cantilever after excitation, where the black dashed line shows an empirical fit of the product of two exponentials which describe the transient perturbation of interest and the cantilever relaxation (see Supplementary Note 1). (d) trEFM map of photoinduced dynamics across a 17.6×10 μm ROI on a sample with the formulation $Cs_{0.17}FA_{0.83}Pb(I_{0.75}Br_{0.25})_3$ collected using 705 nm excitation at an incident intensity of 100 mW/cm$^2$;

Supplementary Fig. 2a shows the topography of this ROI. (e) Correlated PL map of the same ROI, taken using 60× objective, NA=0.7, using a 640 nm laser at a fluence of 105 nJ/cm$^2$. (f) Correlated line traces from the dotted lines shown in (d) and (e), where the trEFM line trace was smoothed to reduce pixel-to-pixel noise. Markers on (f) indicate positions shown in (e) where time-correlated single photon counting histograms were collected and fit with a stretched exponential (see Supplementary Note 2) to extract carrier lifetimes shown in legend (see Supplementary Fig. 2b for additional PL lifetimes collected in this ROI and corresponding correlated surface potential equilibration times).

To further examine this correlation between trEFM equilibration times and electronic carrier dynamics, we investigate the effects of three established surface passivation agents: APTMS,[14–16] AEAPTMS,[14] and PEAI.[54,55] We prepared half-stacks with the architecture glass/ITO/Me-4PACz/$Cs_{0.22}FA_{0.78}Pb(I_{0.85}Br_{0.15})_3$. Supplementary Figs. 3-4 show UV-vis absorbance, XRD, PL, and trPL characterization for the half-stacks. The samples each exhibit Gaussian-shaped PL spectra with a PL maximum at 755 nm, consistent with literature reports for this composition.[58] We fit the trPL decays using stretched exponential functions[36] and summarize the fit parameters in Supplementary Table 1. From the half-stack PL data, we extract an 18×, 2×, and 2× improvement in the carrier lifetime for APTMS, AEAPTMS, and PEAI treatments respectively. Supplementary Fig. 5 show the average PL quantum yields for each half-stack, which show a 13×, 4×, and 2× improvement for APTMS, AEAPTMS, and PEAI respectively. We attribute these enhancements to a reduction in nonradiative recombination pathways.[7,14–17,59]

Fig. 2a shows the surface potential equilibration times we measured with trEFM on untreated and passivated samples plotted against the PL lifetime measured by trPL for those samples. Supplementary Fig. 6 show the trEFM maps for these samples and Supplementary Fig. 4 and Supplementary Table 1 summarize the trPL measurements. We see that the extracted surface potential equilibration time scales proportionally to the carrier lifetime as measured on perovskite half-stacks before and after passivation. These trends are consistent with our interpretation of trEFM probing the time it takes for the surface potential to equilibrate during photoexcitation, largely influenced by the time it takes for carriers to reach new equilibrium profiles. As we reduce nonradiative recombination pathways via passivation, we observe slower surface potential equilibration times that correspond to enhanced minority carrier lifetimes, consistent with general carrier generation and recombination kinetics.[60] However, consistently, the dynamics measured via trEFM are an order of magnitude larger than the average carrier lifetime measured via trPL, indicating that we are not measuring the trPL lifetime with trEFM. This difference is due to the nature of each experiment: trPL probes carrier population decay after pulsed excitation, while trEFM probes the evolution of the carrier population to a new equilibrium distribution in response to a step-function in photoexcitation.

In Fig. 2b, we compare the trEFM time constant with the SRV computed from the trPL data following Wang et al.[13] (see Supplementary Note 3 for additional information on this calculation). We find a strong inverse linear relationship between the dynamics measured by trEFM and the log of the SRV with a Pearson correlation coefficient of -0.91 (with a p-value of 0.013), where a value of -1.0 would indicate a perfect inverse correlation. These results show that the surface potential equilibration time measured by trEFM is predictive of the local SRV, meaning we can use trEFM to characterize treatments for the suppression of nonradiative recombination at perovskite surfaces.

The direct correlation we show between trEFM equilibration times and SRV underpins the opportunity to image carrier dynamics below the optical diffraction limit. Figs. 2c,d show the topography and surface potential equilibration times measured across a 2×1 μm region of an unpassivated perovskite half-stack. We measure surface potential equilibration times that range from 3.7 – 5.6 μs, with grain boundaries exhibiting slower dynamics than grain interiors. Figs. 2e,f show the topography and surface potential equilibration time images for an APTMS-treated sample; APTMS polymerizes on the surface of the sample, changing the observed topography.[15,18,19] As expected from the above analysis, the equilibration times measured by trEFM are indeed significantly (~6×) slower after passivation. Furthermore, we can no longer differentiate grain interiors from grain boundaries. However, we still observe heterogeneous equilibration times across the ROI, indicating that passivated samples still exhibit local heterogeneity in SRV. We show in Supplementary Fig. 7 and Supplementary Table 2 that these results are consistent across different regions and samples.

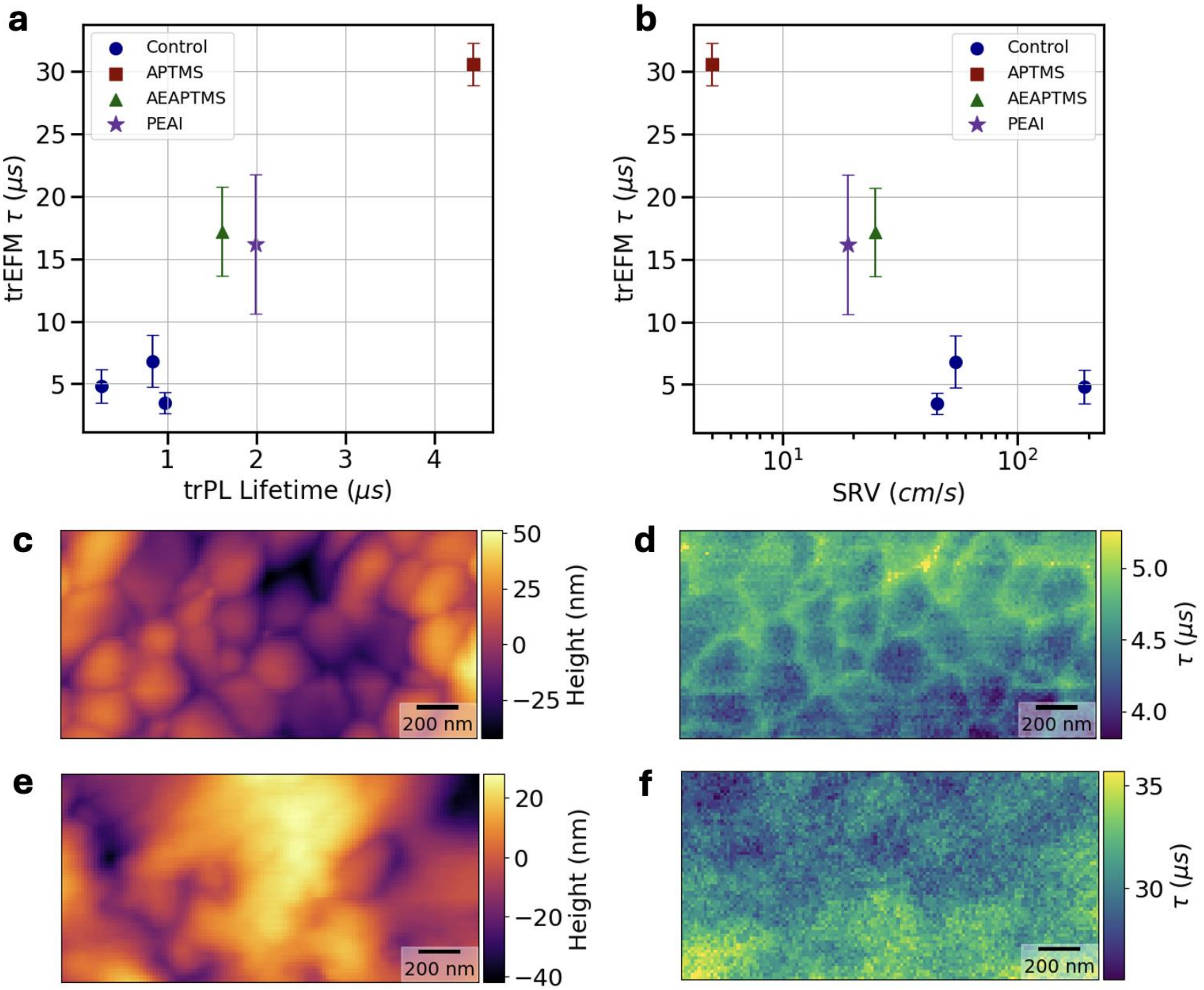

**Fig. 2: Photoinduced dynamics measured by trEFM correlate to PL lifetime and (inversely) to SRV, revealing nanoscale heterogeneity on unpassivated and passivated samples**. (a) Average trEFM τ values plotted against minority carrier lifetimes measured by trPL and (b) approximated SRV (see Supplementary Note 3 for SRV

calculation). All trPL measurements were carried out using a pulsed 640 nm laser operated at ~30 nJ/cm$^2$ – enough to generate $2\times10^{15}$ cm$^{-3}$ carriers (1-2 Suns). (c) Topography of untreated half-stack and (d) corresponding trEFM τ map. (e) Topography of sample after APTMS-passivation with (f) trEFM τ map. All trEFM experiments here used a 405 nm laser with an incident intensity of ~150 mW/cm$^2$ (1.5 Suns) operated in a quasi-steady state regime.

The data so far support our hypothesis that trEFM dynamics probe carrier equilibration during photoexcitation, as dominated by SRV. To further examine this hypothesis, we use the open source 1D drift-diffusion simulator IonMonger, which enables us to explore the effects of various physical parameters, such as mobile ion concentration, excitation wavelength and intensity, and carrier recombination rates, on surface potential dynamics.[61–63] To reproduce our experimental setup, we model a device stack with an inverted, p-i-n architecture and we extract the potential evolution at the active layer surface. We find that the simulated surface potential equilibrates on the microsecond timescale, dependent on both the electronic carrier concentration and mobile ion concentration evolution, consistent with our experimental observations. Supplementary Note 4 describes the relationship between electronic and ionic carrier populations and the surface potential evolution further.

Fig. 3a shows the simulated surface potential evolution as a function of SRV ranging from 0.1 to 1000 cm/s, spanning the range of expected experimental values for untreated and treated films,[12,13,16,36] and Fig. 3b shows the simulated surface potential equilibration time plotted against SRV. As we decrease SRV, the simulated surface potential takes longer to reach the new steady-state, in agreement with the experimental trends from trEFM (Fig. 2b). This result is consistent with general behavior of rate equations and approach to equilibrium, where carrier generation rate is equal to recombination rate: if we decrease the recombination rate by suppressing SRV, then we will reach equilibrium more slowly (see Supplementary Note 4). While we are not attempting to simulate the full 3D system quantitatively, the excellent qualitative agreement between the experimental and simulated timescales further supports our physical interpretation that changes in surface potential equilibration times upon passivation (Fig. 2a,b) are primarily due to suppression of SRV. Supplementary Figs. 8-9 show the relationship between simulated surface potential equilibration time and bulk nonradiative and bimolecular recombination processes. Supplementary Table 3 contains complete simulation parameter details.

We next turn to interpreting the slower trEFM kinetics at the grain boundaries observed in our unpassivated samples. So far, we have established that known reductions in nonradiative recombination result in slower surface potential equilibration times, clearly showing that trEFM dynamics correlate with minority carrier lifetimes and SRVs. However, Fig. 2d shows that defect-rich grain boundaries in unpassivated samples also exhibit relatively slower equilibration times. In our previous work on 2D butylammonium lead iodide ($BA_2PbI_4$) perovskite materials, we also observed slower trEFM dynamics at grain boundaries, which we proposed are caused by increased contributions from slow ionic motion or trap-mediated electronic carrier transport.[31]

We now support this hypothesis with additional simulations of ion motion. Fig. 3c shows the simulated surface potential evolution as a function of the mobile ion concentration ranging from $1\times10^{14}$ cm$^{-3}$ to $1\times10^{17}$ cm$^{-3}$, fixing the ion diffusion coefficient at $1\times10^{-13}$ cm$^2$s$^{-1}$ (parameter details are available in Supplementary Table 3).[64–70] Fig. 3d shows the simulated equilibration times as a function of mobile ion concentration, where higher mobile ion concentrations result in relatively slower surface potential equilibration times. This observation is supported by Poisson's Equation: as we increase the mobile ion concentration, the contribution of slow ion motion to the surface potential dynamics increases, thus lengthening the equilibration times (see Supplementary Note 4 for further discussion).[63] These results suggest that slower surface potential equilibration times measured at grain boundaries indeed result from higher local mobile ion concentrations.

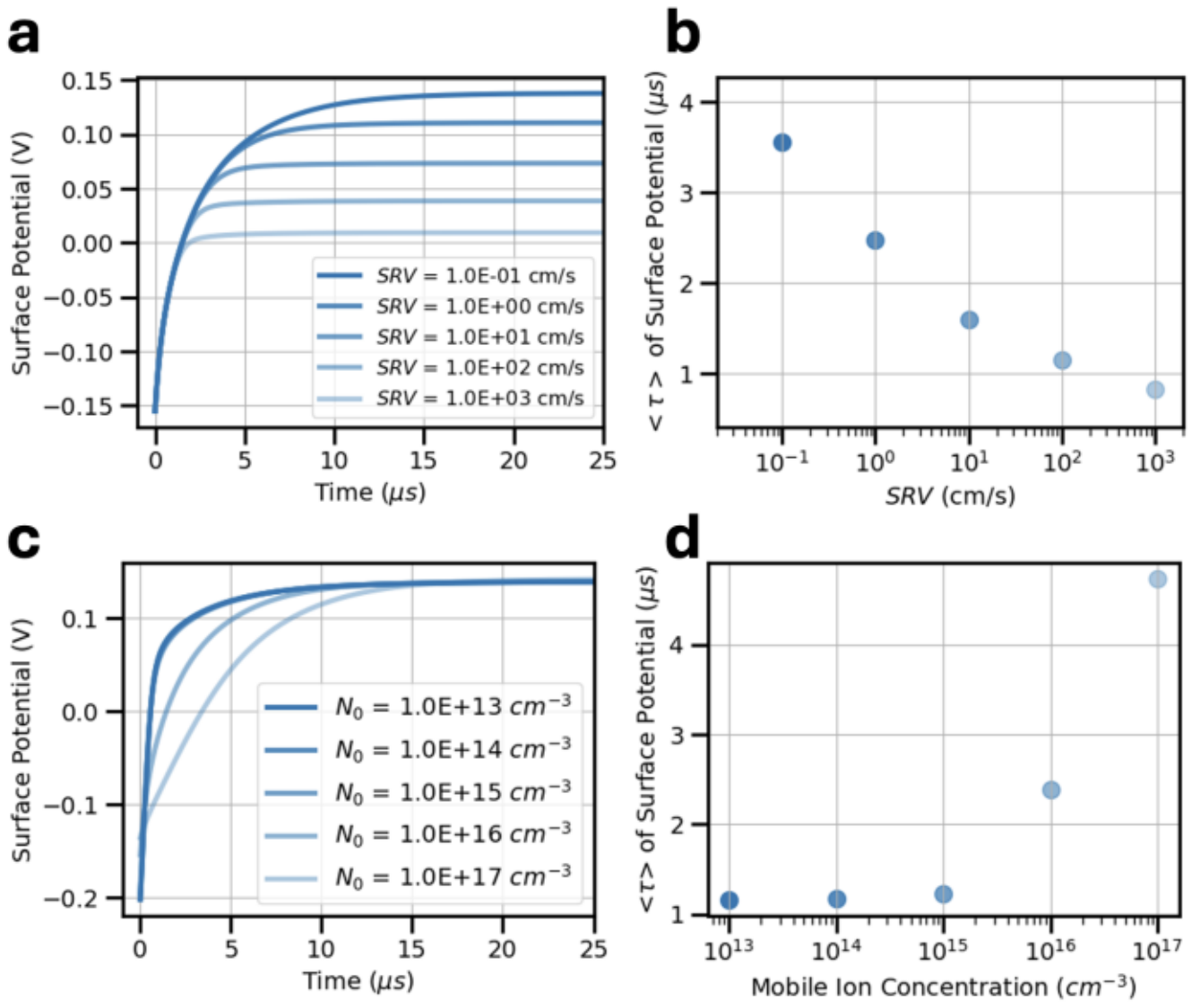

**Fig. 3: Drift-diffusion simulations reveal influence of SRV and mobile ions on surface potential evolution**. (a) Simulated surface potential evolutions plotted against time with varied SRV. (b) The average time constants describing the simulated surface potential equilibration times with respect to SRV, showing slower equilibration behavior at lower SRVs. (c) Simulated surface potential evolutions with varied mobile ion concentrations ($N_0$, $cm^{-3}$). (d) Average time constant that describes the equilibration of the simulated surface potential traces with respect to mobile ion concentration, showing slower equilibration behavior at higher mobile ion concentrations. For complete simulation parameter details see Supplementary Table 3.

To better understand the slower dynamics at grain boundaries, we explore the effect of different background illuminations intensities on the trEFM dynamics. Background illumination creates a population of mobile electronic carriers which can screen charged defects, such as halide vacancies or charge traps, masking their influence on observed dynamics. We thus expect background illumination bias should have two net effects: (1) the surface potential equilibration times should become faster due to overall higher electronic carrier concentrations reducing the marginal contribution from slower ionic processes, and (2) the observed contrast between grain boundaries and grain interiors should homogenize due to screening of both mobile ions and charge traps.

Figure 4 presents data consistent with both these expectations. We apply a continuous illumination bias to the ROI using an LED with a peak wavelength of 660 nm. While under this constant illumination bias, we use the 405 nm laser at 110 $mW/cm^2$ incident intensity to photoexcite the sample for trEFM imaging. We vary the illumination bias intensity from 0 $mW/cm^2$ to 100 $mW/cm^2$. Fig. 4a shows the representative topography of the ROI, with the characteristic nanoscale grains. Fig. 4b-e show the surface potential equilibration time maps measured by trEFM according to increasing illumination bias intensity (where Fig. 4f returns to 0 $mW/cm^2$). Consistent with our hypothesis, Fig. 4 shows that with increasing background illumination intensity, the average equilibration time becomes faster, consistent with our hypothesis above. Fig. 4g shows the surface potential equilibration times for grain boundaries and interiors plotted against illumination bias intensity. We see that with increasing illumination bias, the difference between grain boundaries and interiors decreases, again consistent with our hypothesis (see Supplementary Figs. 10-14 for detailed analysis and replicated results with 660 nm and 455 nm illumination biases). Overall, these results are consistent with our earlier observations and drift-diffusion simulations in both overall equilibration times, and the intensities/photocarrier densities required to mask the grain boundary signals.

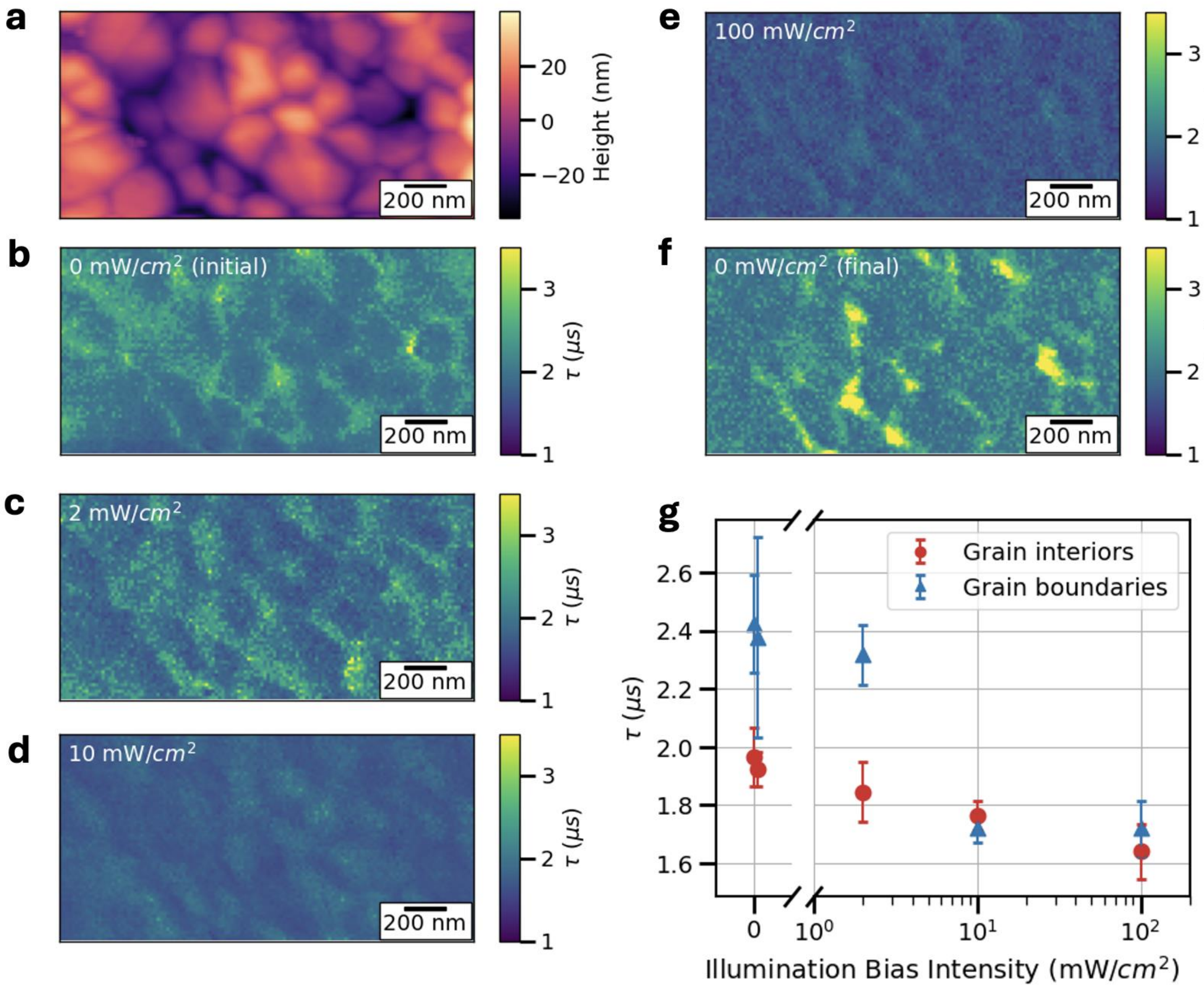

**Fig. 4: Surface potential equilibration with varying background illumination bias intensity**. (a) Topography of 2×1 μm ROI. (b)-(e) Images of surface potential equilibration time constants taken in the same ROI with increasing 660 nm bias illumination intensity ranging from 0 mW/cm$^2$ to 100 mW/cm$^2$, with a step-edge 405 nm laser at 110 mW/cm$^2$ used for trEFM excitation. (f) Reproduced trEFM time constant image at 0 mW/cm$^2$ showing return to unbiased time constants. (g) Average grain interior and boundary trEFM time constants with respect to background illumination bias intensity, error bars show standard deviation of masked pixel selection (Supplementary Fig. 10).

We stress test our model and interpretation with both intensity-dependent and wavelength-dependent trPL and trEFM measurements. Supplementary Figs. 15-19 and Table 4 show the detailed analysis of these experiments and simulations, which are broadly consistent with our interpretation. First, at higher excitation intensities we observe both faster trPL and faster trEFM signatures, consistent with higher order recombination processes and a correspondingly faster approach to equilibrium. Second, since we are exciting the half-stacks through the ITO with the cantilever at the perovskite surface, redder wavelength illumination photoexcites carriers closer to the surface of the film, resulting in a greater contribution from SRV to the surface potential equilibration time.

Putting the experiment and simulation results together, Fig. 5a illustrates the key factors that influence the surface potential dynamics: the local mobile ion concentration and SRV. Fig. 5b shows simulated surface potential evolutions given parameters that approximate unpassivated and passivated grain boundaries and interiors. In the unpassivated regime, slightly higher mobile ion concentrations (grain boundaries) result in slightly slower equilibration times. In the passivated regime, upon the suppression of SRV and reduction in mobile ion concentration[18,19] the equilibration times become slower and indistinguishable. These results suggest that the heterogeneity we observe in surface potential equilibration times after passivation (Fig. 2f) is primarily due to local variations in SRV. This work highlights trEFM's ability to diagnose passivator uniformity and efficiency on the nanoscale.

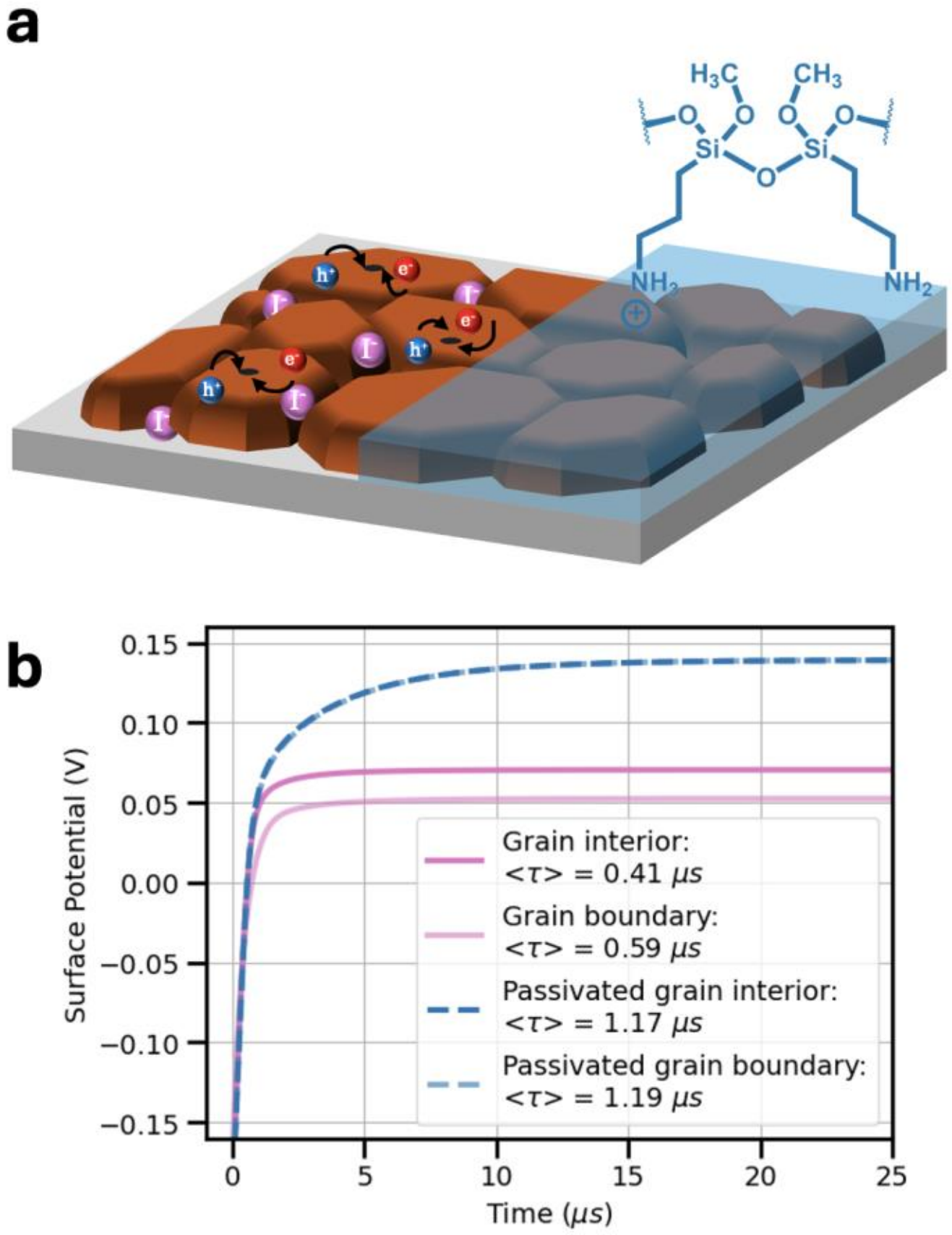

**Fig. 5: Illustration of mobile ions and SRV, which influence surface potential evolution, supported by drift-diffusion simulations**. (a) Illustration describing the effects of SRV and slow ion migration, which are both reduced by surface passivation with APTMS. (b) Drift-diffusion simulations of the surface potential where the primary difference between the grain interior and grain boundary is the concentration of mobile ions ($1\times10^{15}$ cm$^{-3}$ and $5\times10^{15}$

$cm^{-3}$ respectively) with a SRV of 100 cm/s; simulated surface potential evolution of passivated grain interiors and boundaries have both reduced mobile ion concentrations and suppressed SRV values ($5\times10^{14}$ $cm^{-3}$ and $1\times10^{15}$ $cm^{-3}$ for grain interiors and boundaries respectively, with a SRV of 0.1 cm/s). Note that there are two blue, dashed traces representing the passivated condition that are indistinguishable. For complete simulation parameters see Supplementary Table 3.

## Conclusions

We demonstrate that in halide perovskites, trEFM probes local surface potential equilibration time, which is influenced by both electronic carrier recombination and mobile ions. We show that passivation results in slower equilibration times due to strong suppression of SRV, and we directly correlate these equilibration dynamics to SRV obtained from trPL. Importantly, we reveal local variations in SRV after surface passivation using trEFM, demonstrating the need to optimize passivation treatments. Additionally, we conclude that slower equilibration times observed at grain boundaries in unpassivated samples are caused by higher defect concentrations that result in increased contributions of slow ion motion to the surface potential dynamics. These advances highlight the ability to characterize not only the impact of perovskite nanostructure on local carrier dynamics, but also to evaluate the effectiveness of passivation below the optical diffraction limit at length scales relevant to modern halide perovskites. We anticipate the use of sub-diffraction-limited imaging of carrier recombination dynamics to optimize perovskite semiconductor devices and passivation treatments down to the nanoscale.

## Methods

### *Materials*

Lead iodide (99.99%, for perovskite precursor) and Me-4PACz (>99.0%) were purchased from TCI. Lead bromide (99.999%, ultra dry) and cesium iodide (99.999%) were purchased from Alpha Aesar. Formamidinium iodide (>99.99%), phenethylammonium iodide, and methylammonium chloride (>99.99%) were purchased from Greatcell Solar Materials. Aluminum oxide nanoparticles dissolved at 20 wt% was purchased from Sigma Alrich. All other materials were purchased from Sigma Aldrich.

### *Sample preparation*

All preparation of perovskite samples were performed in a $N_2$ glovebox.

A 1.2 M $Cs_{0.17}FA_{0.83}Pb(I_{0.75}Br_{0.25})_3$ solution was prepared according to previous literature reports[57] by mixing the correct molar ratio of $PbI_2$, $PbBr_2$, FAI, and CsI in a 4:1 solvent ratio of N,N-Dimethylformamide (DMF) and Dimethyl sulfoxide (DMSO). 1.5 $cm^2$ indium tin oxide (ITO) coated glass substrates were cleaned by sequential sonication in DI water containing ~2% Micro-90 detergent, DI water, acetone, and isopropanol for 10 minutes each. The substrates were ozone-cleaned for 23 minutes prior to spincoating. Me-4PACz was dissolved in DMF at a concentration of 50 mg/mL and subsequently diluted to 1 mg/mL in isopropanol. Approximately 60 µL of the 1 mg/mL Me-4PACz solution was deposited on the substrate and spincoated at 5000 rpm for 30 s with an acceleration of 800 rpm/s. The film was annealed at 100°C for 10 minutes. Next, 100 µL of a solution with the ratio 1:150 $Al_2O_3$ nanoparticle to isopropanol was spincoated at 6000 rpm for 30 s with an acceleration of 800 rpm/s to wash off excess Me-4PACz and improve wettability.

The substrates were annealed at 100°C for 30 s. Finally, 70 µL of the perovskite precursor solution were deposited on the substrate and spincoated at 1000 rpm for 10 s with an acceleration of 200 rpm/s followed by spinning at 5000 rpm for 35 s with an acceleration of 800 rpm/s; 10 s before the completion of the final spincoating step, 300 µL of anisole was dynamically deposited on the sample. The sample was annealed at 100°C for 45 minutes.

A 1.2 M $Cs_{0.22}FA_{0.78}Pb(I_{0.85}Br_{0.15})_3$ solution was prepared by mixing the correct molar ratio of $PbI_2$, $PbBr_2$, FAI, CsI and a 15 mol% relative to the Pb content equivalent of MACl to improve crystallization in a 4:1 solvent ratio of N,N-Dimethylformamide (DMF) and Dimethyl sulfoxide (DMSO). Indium tin oxide (ITO) coated glass substrates 1.5 $cm^2$ in size were cleaned by sequential sonication in DI water containing ~2% Micro-90 detergent, DI water, acetone, and isopropanol for 10 minutes each. The substrates were ozone-cleaned for 23 minutes prior to spincoating. Me-4PACz was first dissolved in DMF at a concentration of 50 mg/mL and was subsequently diluted to 1 mg/mL in 2-propanol (IPA). Approximately 60 µL of the 1 mg/mL Me-4PACz solution was spincoated on the substrates at 5000 rpms for 30 s with an acceleration of 800 rpm/s and annealed at 100 °C for 10 min. Next, 100 µL of IPA was dynamically spincoated on the substrate to wash away excess Me-4PACz at 6000 rpm for 30 s with an acceleration of 800 rpm/s and annealed at 100 °C for 5 min. Following that, $Al_2O_3$ nanoparticles suspended in IPA was diluted further with IPA by 1:150. Approximately 50 µL of the $Al_2O_3$ solution was spincoated at 6000 rpm for 30 s and annealed at 100 °C for 50 s to improve the wettability of perovskite on the Me-4PACz layer. The perovskite solution was filtered through a 0.2 µm PTFE membrane filter. Approximately 60 µL of the perovskite solution was dynamically deposited on top of the substrate and spincoated at 1000 rpm for 10 s with an acceleration of 200 rpm/s followed by spinning at 5000 rpm for 35 s with an acceleration of 800 rpm/s. After 25-30 s of the second step, 330 µL of an anisole antisolvent was dynamically deposited on the spinning substrate. The half-stack was then annealed for 30 min at 100 °C on a hot plate.

***Passivation***

APTMS and AEAPTMS passivation of the perovskite half-stacks was performed at room temperature and ambient conditions in a vacuum oven for 4 mins. A volume of 500 µL of the silane was placed in a 4 mL vial with the perovskite sample placed face up near the vial. The vial and sample were covered with a 500 mL glass jar inside of the vacuum chamber, and the pressure was pumped down to -26 inHg relative to the atmospheric pressure, as described in previous work.[16]

PEAI passivation was performed in a nitrogen glovebox by first preparing a 0.030 M solution of PEAI in IPA. Next, 100 µL of the PEAI solution was dynamically spincoated on the perovskite surface at 3000 rpm for 30 s with an acceleration of 800 rpm/s. Following treatment with PEAI, the sample was left to rest in the nitrogen glovebox in the dark for ~24 hours prior to making measurements.[55]

***UV-Vis characterization***

UV-Vis absorption spectra of the perovskite control and passivated half-stacks were collected using an Agilent 8453 UV-Vis Spectrometer in a wavelength range of 200-1100 nm and an integration time of 0.5 s.

***XRD characterization***

X-ray diffraction measurements of the perovskite control and passivated half-stacks were measured using a Bruker D8 Discover with a Pilatus 100 K large-area 2D detector with Cu Kα radiation at 50,000 mW.

***Photoluminescence characterization***

trPL was measured using a PicoQuant Fluotime 100 spectrometer and Picoharp 300 TCSPC system equipped with a 640 nm pulsed diode laser with a high average power of 30 mW and 90 ps pulse width. The repetition rate of the pulsed laser was controlled with an external function generator. The laser was pulsed with an excitation intensity of ~ 30 nJ/cm$^2$, in order to capture the decay at a carrier density near that of one Sun power. The PL emission was passed through a 700 nm long-pass filter before reaching the detector. The PL data was fitted using a stretched exponential function, additional information of this fitting can be found in the Supporting Information. For this measurement, the samples were measured in ambient conditions immediately after removal from a nitrogen glovebox.

Steady-state PL spectra and PLQY values were collected using an integrating sphere spectrometer (Hamamatsu Photonics K.K.) and a 532 nm continuous wave laser (CrystalLaser Lc). The integration time of the spectrometer was set to 250 ms and 5 measurements were averaged for each spectrum. The laser fluences was measured with a Thorlabs beam profiler (BP209-VIS). A continuous neutral density filter wheel (Thorlabs) was used to control the laser fluence. The neutral density filter wheel was adjusted for all measurements to be made at 100 mW/cm$^2$.

***Confocal photoluminescence characterization***

Confocal PL imaging was performed using a custom scanning confocal microscope built with a Nikon TE-2000 inverted microscope fitted with 60x dry objective (Nikon Plan Fluor, NA=0.7). A 640 nm pulsed diode laser (PDL 800-D P-C-640B, FWHM = 90 ps) was used at a repetition rate of 250 kHz, with a resulting fluence of 105 nJ/cm$^2$. The laser was coupled to the microscope via a FC/APC single-mode fiber and directed into the objective via a 640 dichroic cube. Sample was encapsulated under nitrogen and excited face-on (not through the substrate). The emission was filtered through a 700-850 nm bandpass filter (700 LP and 850 SP) and focused on a Single Photon Avalanche Diode from Micro Photon Devices. The PL decay was collected using a PicoHarp 300 time-correlated single photon counting (TCSPC) module. The scanning is done using a piezo nano position stage (PI). The data collection and analysis are performed through custom Python-based software.

***trEFM characterization***

trEFM and the required experimental setup have been discussed in detail.[48,49] Measurements were performed on an Asylum Research MFP3D-BIO mounted on a Nikon inverted optical microscope. All measurements used Pt-coated cantilevers (mikroMasch HQ/NSC15/Pt) driven at resonance frequency (~300 kHz). The sample was mounted in an inert glovebox environment in a sealed cell, then imaged under active flowing nitrogen in this sealed cell. The cantilever oscillation was recorded using a 16-bit A/D digitizer (Dynamic Signals/GaGe Razor Express CSE1622), typically at 5 MS/s, and synced to the cantilever oscillation phase (180°) using custom trigger electronics (detailed circuit information can be found in our previous reports; additional information available upon request).[48,49] In an experimental window of 16 ms, we apply a bias of +10 V to the cantilever at t = 1 ms, the sample is then allowed to equilibrate for 4 ms; the cantilever oscillation is digitized starting at t = 4.6 ms through t = 8.6 ms, and laser excitation is triggered at t = 5 ms and turned off at t = 7 ms. We used either a 405 nm (Omicron PhoxXplus 405-120) or a 705 nm (Omicron PhoxXplus 705-40) continuous wavelength laser at an incident intensity of ~110-150 mW/cm$^2$ to excite our samples; we adjust the intensity using neutral density filters and the electrical power via software control. The laser was focused via a bottom objective on an inverted optical microscope and co-aligned with the cantilever tip. The illumination intensity was measured using the combination of a calibrated photodiode and a Pixera CCD camera (150CL-CU). Raw cantilever deflection data were used to extract the instantaneous frequency by demodulating the time-dependent cantilever amplitude using a short-time Fourier transform (STFT) and then using a parabolic estimation of the peak frequency (ridge finding) per time segment. This code is freely available online via the FFTA Python package (https://github.com/GingerLabUW/FFTA).

**Author Contributions**

M.D.B. and J.P. contributed equally to this work. M.D.B. and J.P. wrote the paper and prepared the figures. M.D.B., J.P., R.G., and D.S.G. conceptualized the work. M.D.B. and J.P. drafted and edited the manuscript, with input and revisions from R.G. and D.S.G. Finally, M.D.B. and J.P. performed all the experiments.

**Acknowledgements**

The atomic force microscopy imaging work (M.D.B., J.P., R.G.) was supported by the U.S. Department of Energy, Office of Basic Energy Sciences, Division of Materials Sciences and Engineering under Award DE-SC0013957. Part of this work was carried out at the Molecular Analysis Facility, a National Nanotechnology Coordinated Infrastructure site at the University of Washington, which is supported by the National Science Foundation (NNCI-1542101), the Molecular Engineering and Sciences Institute, and the Clean Energy Institute. M.D.B. acknowledges support from the NSF Graduate Student Fellowship program under Grant no. DGE-2140004. M.D.B. thanks Nir Tessler for advice on drift-diffusion simulations. J.P. thanks Zixu Huang for help with XRD measurements. D.S.G acknowledges salary and infrastructure support from the Washington Research Foundation, the Alvin L. and Verla R. Kwiram endowment, and the B. Seymour Rabinovitch Endowment.

# Supplementary Information

# Sub-diffraction-limited Imaging of Carrier Dynamics in Halide Perovskite Semiconductors: Effects of Passivation, Morphology, and Ion Motion

*Madeleine D. Breshears[a,+], Justin Pothoof[a,+], Rajiv Giridharagopal[a], David S. Ginger[a,*]*

[a]Department of Chemistry, University of Washington, Box 351700, Seattle, Washington, 98195-1700, United States

[*]Corresponding Author: dginger@uw.edu

[+]M.D. Breshears and J. Pothoof contributed equally to this work.

# Table of Contents

**Supplementary Note 1.** Cantilever calibration for trEFM time constant extraction.

The experimental and theoretical background behind trEFM is described in significant detail in our previous works[1,2] and by Tirmzi, et al.[3] Here, we summarize the cantilever calibration procedure for the sake of reference.

During the trEFM experiment, we digitize the deflection or the raw oscillation of the cantilever, and then we demodulate it using a Short-Time Fourier Transform (STFT). Many demodulation methods work, such as non-stationary Fourier mode decomposition,[4] which demonstrates improved signal-to-noise extraction but is computationally expensive and slow. Hilbert Transforms[1,5] are computationally fast, but less robust to the experimental noise we observe. Thus, we use STFT to demodulate our data to balance the signal-to-noise we obtain and computation time. We then use a product of two exponentials (**Equation 1**) to empirically fit the change in frequency (Δf), including the transient response and cantilever relaxation (shown in main text Fig. 1c), and extract the time it takes for the cantilever's frequency to shift to its maximum deviation from the drive frequency after the illumination trigger.

**Equation 1**.

$$\Delta f = A \times \left[\exp\left(-\frac{t}{\tau_1}\right) - 1\right] \times \left[-\exp\left(-\frac{t}{\tau_2}\right)\right]$$

Where A is a coefficient that accounts for the magnitude of the frequency shift, t is time, $\tau_1$ describes the transient decay, and $\tau_2$ describes the cantilever relaxation.

We call the time it takes for the cantilever's frequency to shift to its maximum deviation from the drive frequency the "time-to-first-peak" or $t_{fp}$. The $t_{fp}$ is cantilever-dependent, meaning physical qualities of the cantilever like its spring constant or quality factor affect the frequency response. Because the frequency response is cantilever-specific, we must extract the cantilever-independent dynamics via a calibration of the frequency response to a time constant that describes the transient response of interest (main text Fig. 1c). We collect the relevant cantilever parameters during the trEFM experiment, enabling us to use our model[6] to simulate a $t_{fp}$ to τ calibration curve (Supplementary Fig. 1), which we use to obtain the cantilever-independent equilibration time maps shown in this work.

**Supplementary Fig 1.** Example simulated calibration curve for extraction of cantilever-independent τ values.

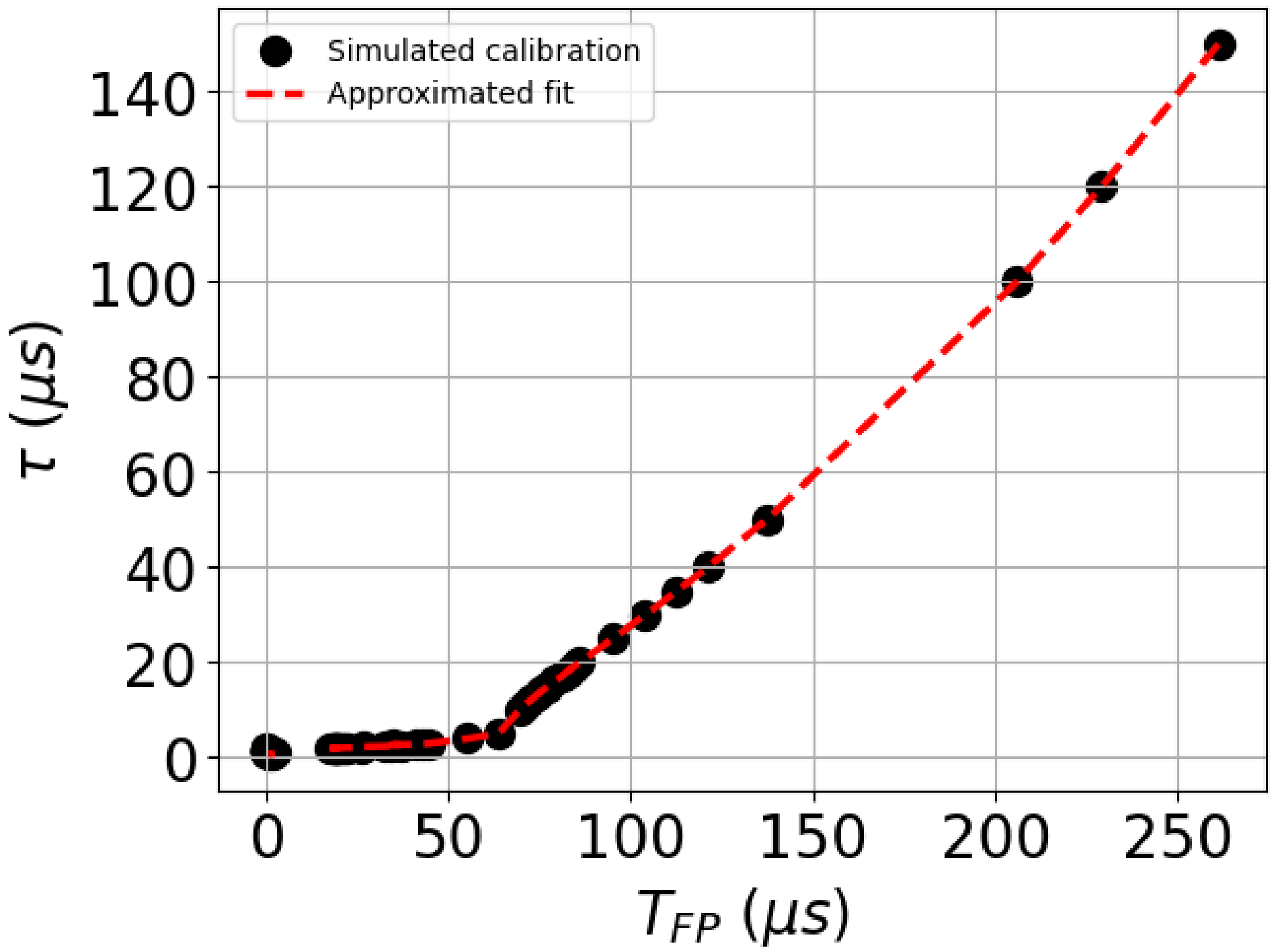

Figure shows characteristic time constant, τ, plotted against time-to-first peak, $t_{fp}$, obtained from simulation; this calibration curve is used to map experimentally obtained $t_{fp}$ values to cantilever-independent τ values that describe the external force perturbation to the cantilever's oscillation (refer to Supplementary Note 1). We use our publicly available FFTA Python package to generate this calibration curve.[6]

**Supplementary Fig 2.** Topography of large-scale ROI to correlate trEFM with confocal PL microscopy.

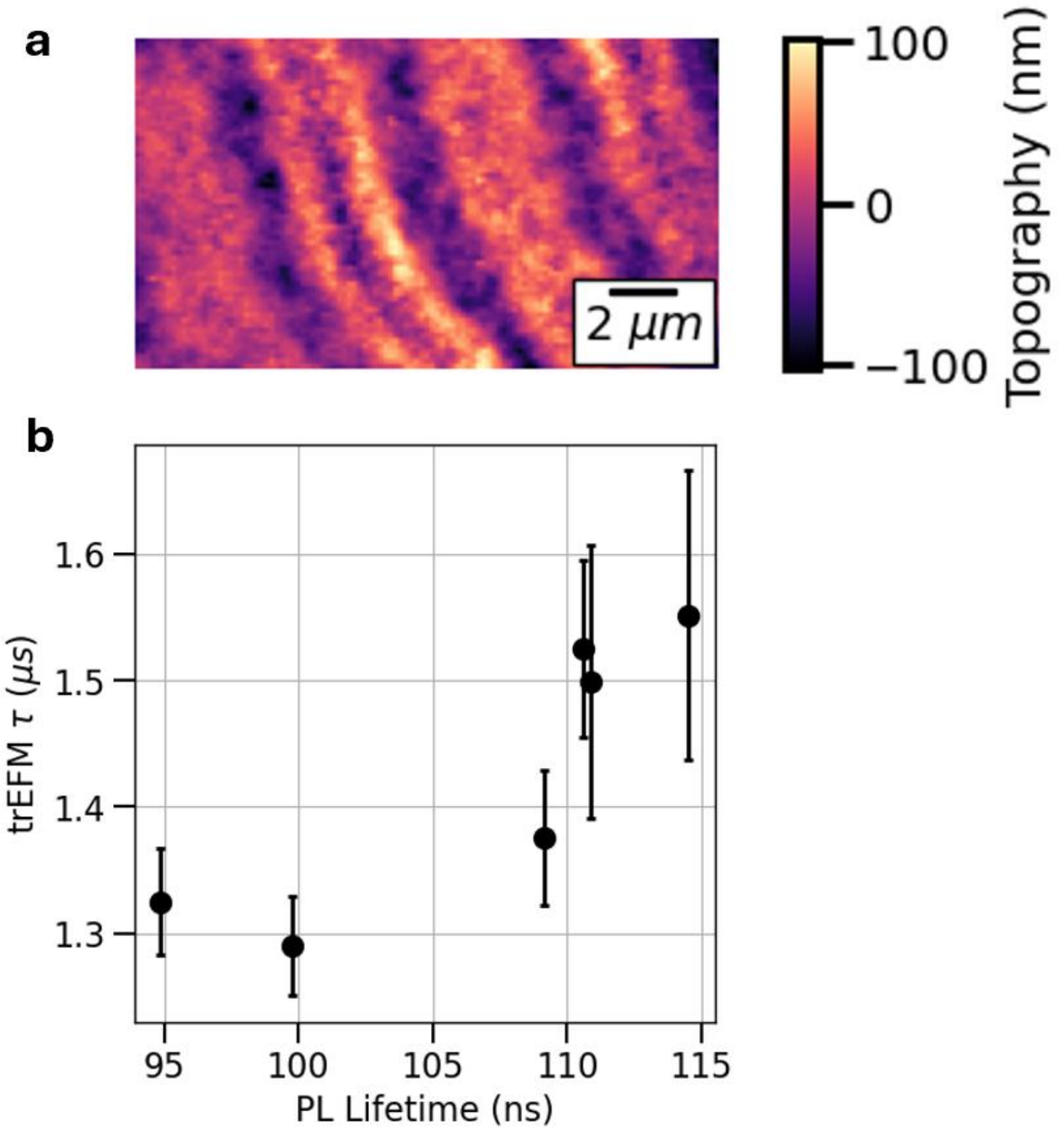

**(a)** Topography of large-scale ROI shown in main text Fig. 1d-e. **(b)** trEFM surface potential equilibration time constants plotted against trPL lifetimes in correlated regions. trPL lifetimes were collected with using 60× objective, NA=0.7, using a 640 nm laser at a fluence of 105 nJ/cm$^2$.

**Supplementary Note 2.** Fitting trPL decays with stretch exponential.

We fit the measured trPL decays using a stretched exponential function as described in **Equation 2** below. The stretched exponential function uses both a characteristic lifetime ($\tau_C$) and a β-factor which describe the time to reach 1/e of the initial signal and the degree of heterogeneity in emitting states; a β-factor closer to 1 means emission is more homogenous, and the stretched exponential function collapses into a mono-exponential function.[7]

**Equation 2.**

$$y = \exp\left(-\frac{t}{\tau_C}\right)^{\beta}$$

From the $\tau_C$ and β-factor, we calculated the average trPL lifetimes as shown in **Equation 3**; this takes into account the amount of heterogeneity that exists in halide perovskite emission.[7]

**Equation 3.**

$$< \tau > = \frac{\tau_C}{\beta}\Gamma\left(\frac{1}{\beta}\right)$$

**Supplementary Fig 3.** UV-vis absorbance and XRD spectra to characterize perovskite half-stack samples and passivations.

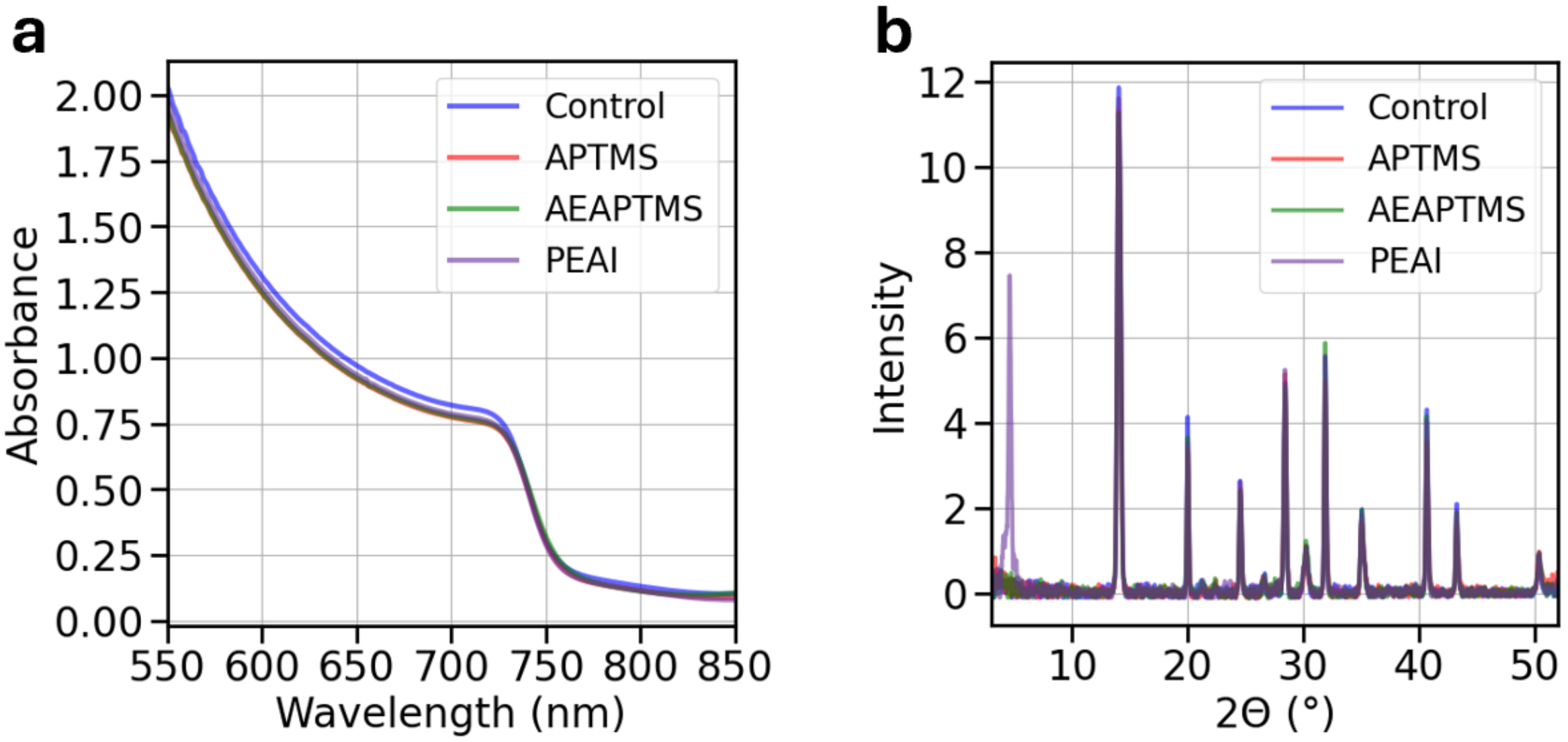

**(a)** UV-vis absorbance of half-stack samples including the unpassivated control and APTMS-, AEAPTMS-, and PEAI-passivated samples showing negligible change in overall absorbance after passivation. UV-vis characterization protocol described in main text Methods. **(b)** XRD spectra of half-stack samples including the unpassivated control and APTMS-, AEAPTMS-, and PEAI-passivated samples showing negligible change in perovskite diffraction peaks and presence of PEAI salt peak at 4.6°.[8]

**Supplementary Fig 4.** PL spectra and trPL decays of perovskite half-stacks before and after surface passivation treatments.

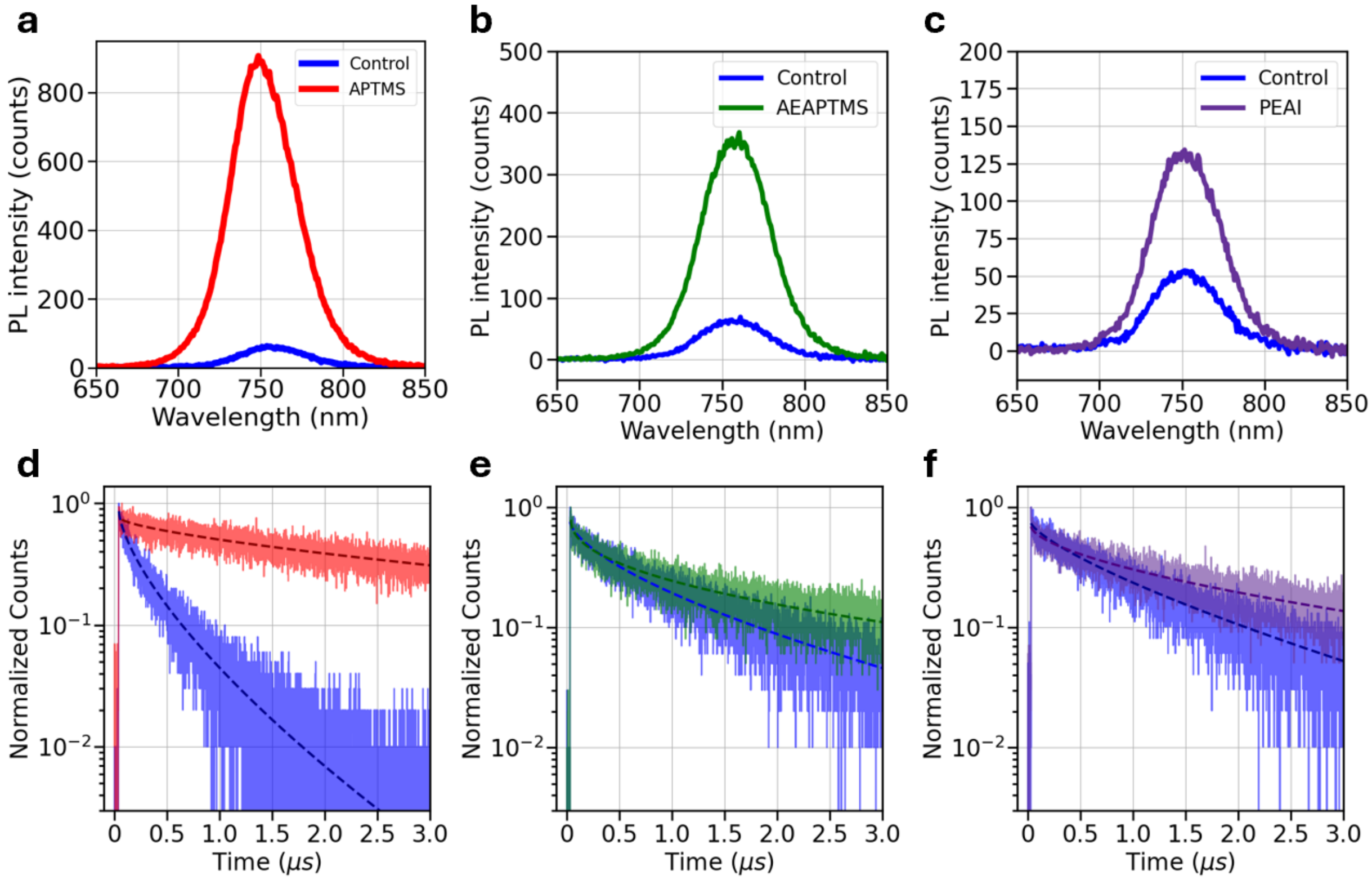

**(a)-(c)** PL spectra of control and passivated half-stack samples for APTMS, AEAPTMS, and PEAI passivation respectively. **(d)-(f)** trPL decays for control and passivated half-stack samples for APTMS, AEAPTMS, and PEAI passivation respectively.

**Supplementary Table 1.** Stretched exponential fitting parameters for Glass/ITO/Me-4PACz/$Cs_{0.22}FA_{0.78}Pb(I_{0.85}Br_{0.15})_3$ half-stack samples before and after passivation with AEAPTMS and PEAI.

| Sample | $\tau_C$ (ns) | β-factor |
|---|---|---|
| **Control** | 598 | 0.64 |
| **AEPTMS** | 684 | 0.46 |
| **Control** | 814 | 0.75 |
| **PEAI** | 1335 | 0.60 |

**Supplementary Fig 5**. Photoluminescence quantum yields (PLQY) for control and passivated samples.

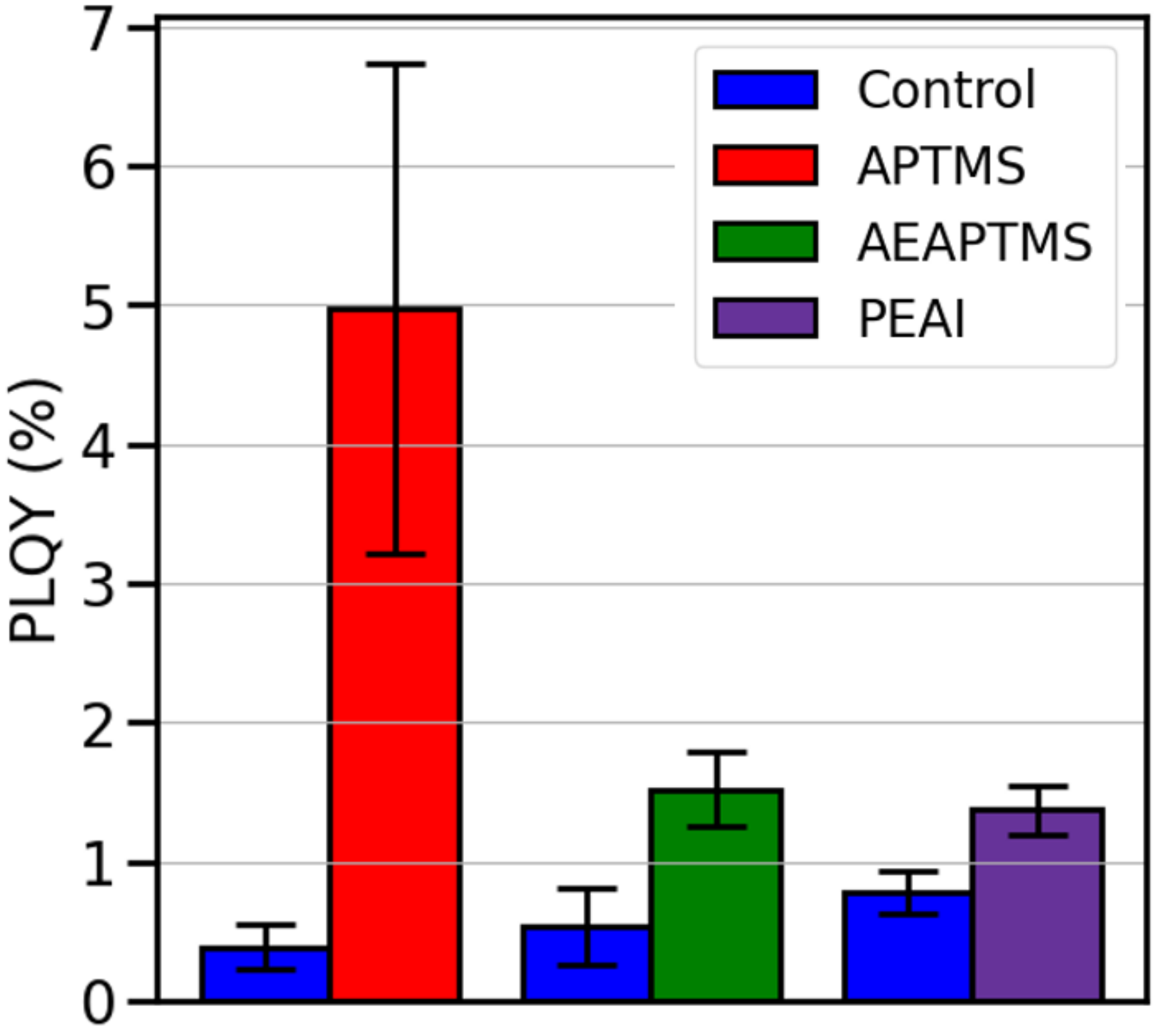

Figure shows PLQY enhancement after passivation with each passivator. The bar represents the average of 3 different samples and the error bars are the standard deviation.

**Supplementary Fig 6.** trEFM measurements of AEAPTMS and PEAI treatment.

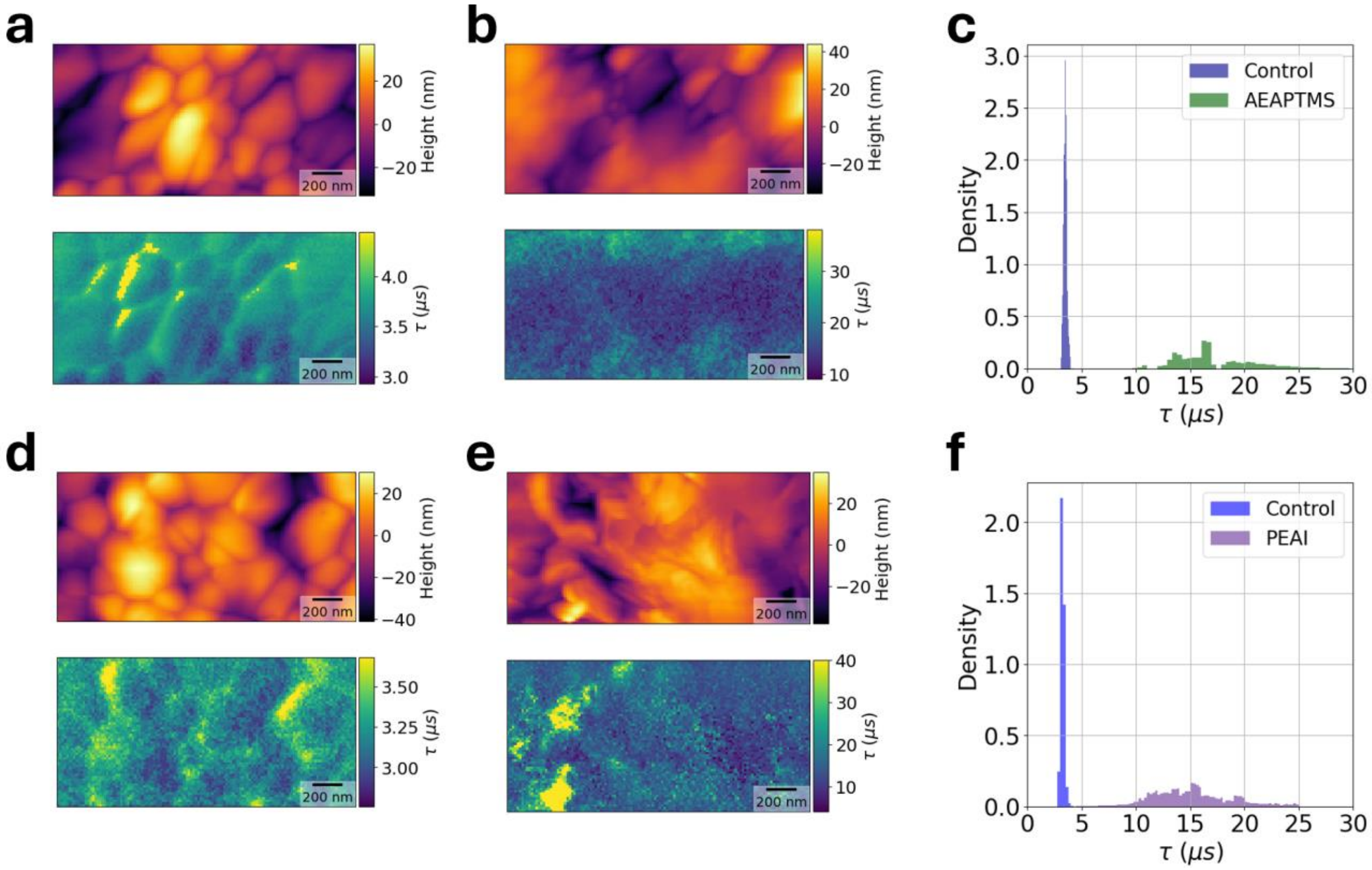

**(a)** Representative topography and trEFM time constant image of control and **(b)** AEAPTMS treated half-stack samples. **(c)** Associated histogram that shows slower surface potential equilibration times after treatment. **(d)** Representative topography and trEFM time constant image of control and **(e)** PEAI treated samples. **(f)** Associated histogram that also displays slower surface potential equilibration times after treatment.

**Supplementary Note 3.** Approximation of surface recombination velocity (SRV) from trPL effective lifetimes.

We approximate the SRV at the surface of the perovskite from the trPL measurements using **Equation 4**. Here, $\tau_{surf}$, W, and D describe the carrier lifetime at the surface, the perovskite thickness (500 nm), and electronic carrier diffusion coefficient (assumed to be 0.75 $cm^2/s$). We can determine $\tau_{surf}$ from the effective carrier lifetime ($\tau_{eff}$) and the bulk carrier lifetime ($\tau_{bulk}$) as shown in **Equation 5**. We take the average trPL lifetimes as the $\tau_{eff}$ and assume the $\tau_{bulk}$ to be 8000 ns and the surface recombination velocity at the perovskite/substrate interface to be negligible, as previous reported.[9,10]

**Equation 4.**

$$SRV = \frac{W}{\tau_{surf} - \left(\frac{4}{D}\right)\left(\frac{W}{\pi}\right)^2}$$

**Equation 5.**

$$\tau_{surf} = \left(\frac{1}{\tau_{eff}} - \frac{1}{\tau_{bulk}}\right)^{-1}$$

**Supplementary Fig 7.** Histograms of the trEFM surface potential equilibration times measured in three ROIs of a control and APTMS-treated $Cs_{0.22}FA_{0.78}Pb(I_{0.85}Br_{0.15})_3$ half-stack sample.

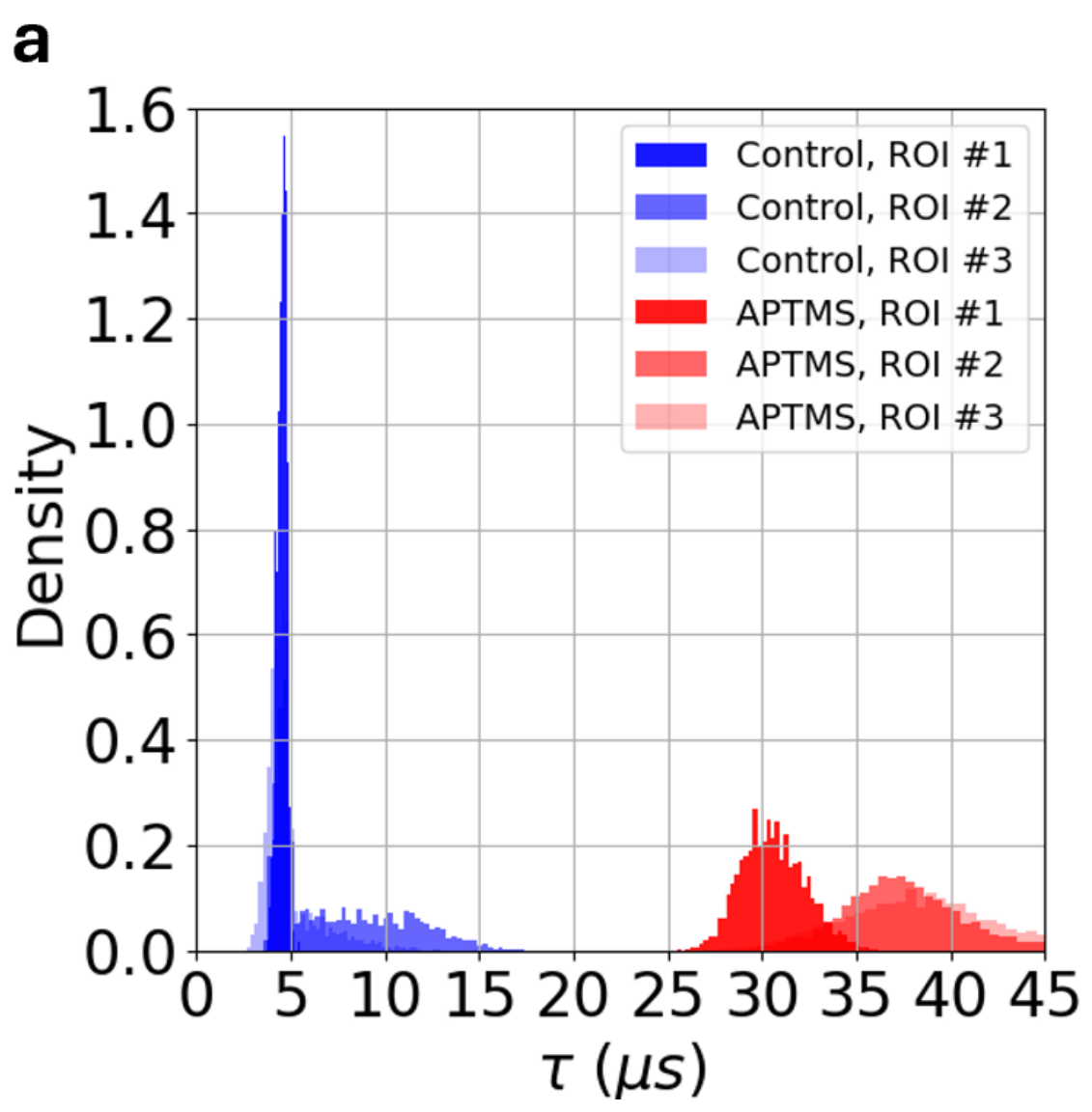

**(a)** Density normalized histograms showing the trEFM time constants measured across three unique ROIs in a Glass/ITO/Me-4PACz/$Cs_{0.22}FA_{0.78}Pb(I_{0.85}Br_{0.15})_3$ half-stack before and after APTMS treatment. The control and APTMS ROIs are also unique from one another.

**Supplementary Table 2.** Average values of the trEFM surface potential equilibration times measured in three separate $Cs_{0.22}FA_{0.78}Pb(I_{0.85}Br_{0.15})_3$ half-stack samples before and after APTMS passivation.

| Sample | Control <τ> (μs) | APTMS <τ> (μs) |
|---|---|---|
| **1** | 10.5 | 35.0 |
| **2** | 5.4 | 14.2 |
| **3** | 5.5 | 35.5 |

**Supplementary Note 4:** On surface potential equilibration in perovskite systems as modeled by drift diffusion simulations.

We assume that the surface potential in our simulations is dominated by electronic carrier dynamics because we observe equilibration of the surface potential on the scale of microseconds, while ion migration is typically on the minute timescale (see **Equation 6**).[11,12]

**Equation 6**.

$$\tau_{ion} = \frac{b}{D_I}\sqrt{\frac{V_T \varepsilon_A}{q\widehat{N}_0}}$$

Where $\tau_{ion}$ is the time constant that describes mobile ions migrating to the Debye layers in perovskite thin films, b is the sample thickness, $D_I$ is the ion diffusion coefficient, $V_T$ is the built in voltage, $\varepsilon_A$ is the dielectric constant of the absorber, q is the elementary charge, and $\widehat{N}_0$ is the mean cation vacancy density (cation vacancies are assumed to be relatively stationary compared to anion vacancies, and cation vacancy density is assumed to be equal to mean anion vacancy density, so $\widehat{N}_0$ represents mobile ion concentration).[11–13] Given the parameters in **Supplementary Table 3**, ion motion occurs on the scale of several minutes. Because the electric potential in the sample is calculated according to Poisson's Equation ($\frac{d^2\phi}{dt^2} = \frac{q}{\varepsilon_A}\left(\widehat{N}_0 - P + n - p\right)$, where ϕ is potential, P is mobile anion vacancy, n is electron concentration, and p is hole concentration), we can use electronic carrier dynamics to explain the surface potential equilibration timescale. That said, if we increase the concentration of mobile ions (increase $\widehat{N}_0$, because $\widehat{N}_0$= P), we will observe slower surface potential equilibration dynamics due to the proportionally larger contribution from extremely slow ion motion.

We propose that the carrier equilibrium condition can be described by the following set of equations:

**Equation 7**.

$$G = R = -\frac{dn}{dt}$$

Where G is carrier generation rate, R is carrier recombination rate, n is electronic carrier concentration, and t is time. We describe recombination with the following equation:

**Equation 8**.

$$R = \frac{1}{\tau_{SRH}}[(n_0 + \Delta n)] + \beta[(n_0 + \Delta n)(p_0 + \Delta p)]$$

Where $\tau_{SRH}$ is the nonradiative recombination carrier lifetime, $n_0$ and $p_0$ are the electron and hole concentrations in the dark, Δn and Δp are the change in electron and hole concentrations, and β is the bimolecular recombination rate constant. Here, we assume that there is no Auger-Meitner recombination.[7,11] At the surface, there is an additional recombination term to account for surface recombination velocity, which we describe with the following:

**Equation 9.**

$$R_{surface} = SRV \times \Delta n$$

Where SRV is the surface recombination velocity at the perovskite/surface interface. Given our electronic carrier equilibrium condition described in **Equation 7** (G = R), carrier concentrations in a sample with a higher SRV term will reach equilibrium at earlier times than a sample with lower SRV.

We fit the surface potential evolution with a tri-exponential function to extract an average time constant to describe the rise time, though we note that this time constant is a phenomenological representation of surface potential equilibration dynamics and the trends should be interpreted as qualitative rather than quantitative in capturing the trEFM experiment.

**Supplementary Fig 8.** Effects of the nonradiative recombination rate constant on the surface potential equilibration time.

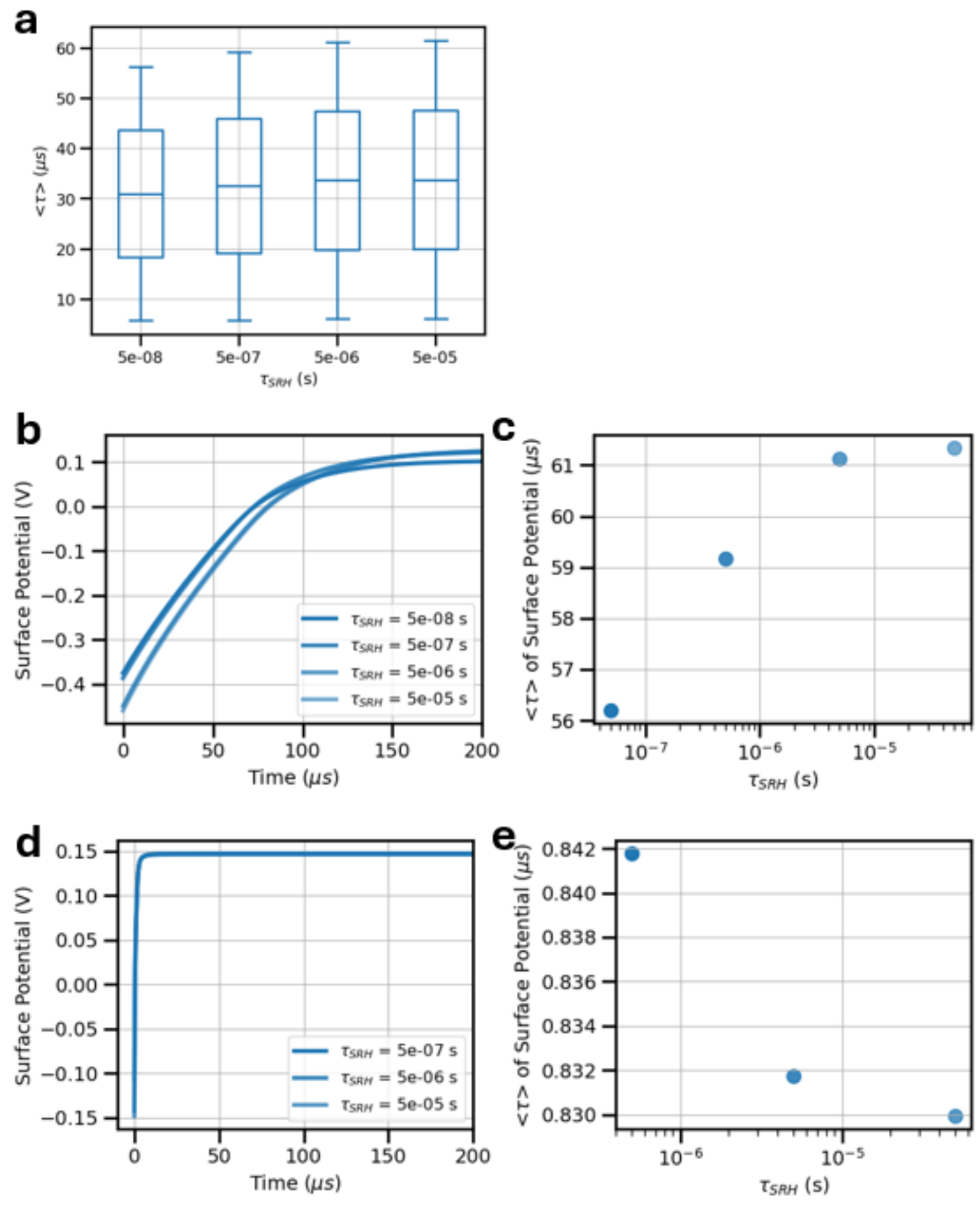

**(a)** Boxplot showing <τ> values calculated from tri-exponential fits of simulated surface potential traces with 405 nm, 1 mW/cm$^2$ excitation, showing slight increase in surface potential equilibration time with increasing Shockley Read Hall carrier lifetime ($\tau_{SRH}$); at low fluences, increasing nonradiative recombination lifetime leads to slower equilibration times, which is also shown in **(b)** example surface potential evolutions and **(c)** <τ> fits for those examples; however, at higher excitation intensities like 100 mW/cm$^2$, $\tau_{SRH}$ shows little to no effect on the surface potential equilibration time, as seen in **(d)** and **(e)**. Because $\tau_{SRH}$ primarily affects bulk carrier dynamics, we expect that the surface recombination velocity dominates the carrier dynamics at the surface of the absorber. For full simulation details, see Supplementary Table 3, where the primary parameters swept for **(b)-(e)** are $\tau_{SRH}$ (s) and illumination intensity (scalar), and all other parameters are held constant. Boxplot mid-line is the median, box edges are the first and third quartiles of the data, and the whiskers show the range of the data.

**Supplementary Fig 9.** Effects of the bimolecular recombination rate constant on the surface potential equilibration time.

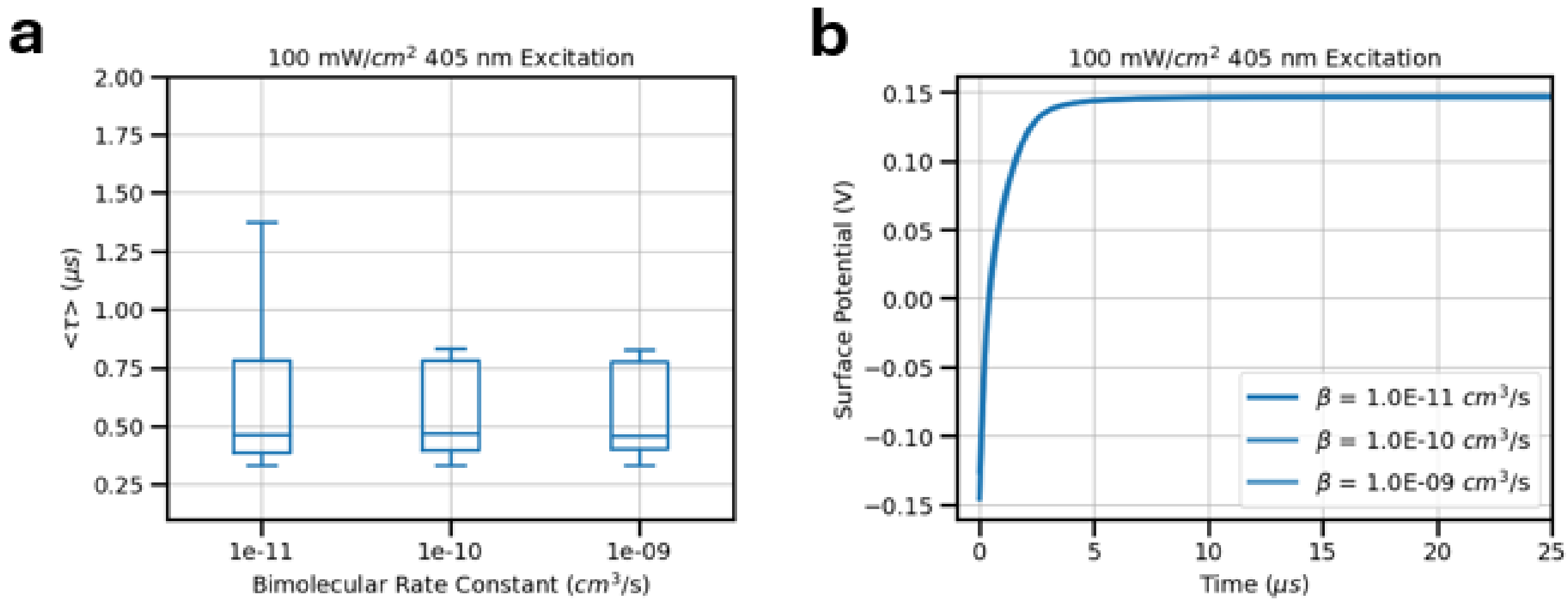

**(a)** Boxplot showing <τ> values calculated from a tri-exponential fit of simulated surface potential versus bimolecular rate constants, where simulated excitation was 405 nm at 100 mW/$cm^2$, similar to experimental conditions described in text; **(b)** example surface potential traces with varied bimolecular rate constants for 100 mW/$cm^2$ 405 nm excitation with all other parameters held constant. For full simulation details, see Supplementary Table 3. We recognize that bimolecular recombination rate may be affected by surface passivation,[14] but these simulations show that the bimolecular rate constant has a relatively minor effect on the surface potential evolution. Boxplot mid-line is the median, box edges are the first and third quartiles of the data, and the whiskers show the range of the data.

**Supplementary Table 3.** IonMonger simulation parameters used to simulate perovskite surface potential evolution.

| Parameter | Value(s) | Reference |
|---|---|---|
| **Thickness** | 500 nm | N/A |
| **Bandgap** | 1.7 eV | [11] |
| **Absorption coefficient (α)** | 405 nm: $2.42085\times10^{7}$ m$^{-1}$<br>705 nm: $2.5807\times10^{6}$ m$^{-1}$ | [7] |
| **Photon flux ($F_{ph}$)** | 405 nm: $2.05\times10^{21}$ m$^{-2}$s$^{-1}$<br>705 nm: $3.55\times10^{21}$ m$^{-2}$s$^{-1}$ | N/A |
| **Illumination Intensity (scalar)** | 0.01 – 10 (scalar of 1 used for all 100 mW/cm$^{2}$ simulations) | N/A |
| **Dielectric constant ($\varepsilon_A$)** | $24.1\times\varepsilon_0$ Fm$^{-1}$ | [11,15,16] |
| **Mobile ion concentration ($N_0$)** | $1\times10^{13}$ - $1\times10^{17}$ cm$^{-3}$ | [17–23] |
| **Ion diffusion coefficient ($D_I$)** | $1\times10^{-13}$ cm$^{2}$s$^{-1}$ | [16–23] |
| **Carrier diffusion coefficient (D)** | $7.5\times10^{-3}$ – $7.5\times10^{-1}$ cm$^{2}$s$^{-1}$ | [7,16,24,25] |
| **Nonradiative recombination lifetime ($\tau_{SRH}$)** | $5\times10^{-8}$ - $5\times10^{-5}$ s | [7,24] |
| **Bimolecular recombination rate constant (β)** | $1\times10^{-11}$ - $1\times10^{-9}$ cm$^{3}$s$^{-1}$ | [7,11,15] |
| **Auger-Meitner recombination rate** | $1\times10^{-28}$ cm$^{6}$s$^{-1}$ | [7,15] |
| **Surface recombination velocity (SRV)** | $1\times10^{-1}$ - $1\times10^{3}$ cm s$^{-1}$ | [7,9–11,15,26] |

**Supplementary Fig 10.** Binary mask used to separate grain interiors and boundaries.

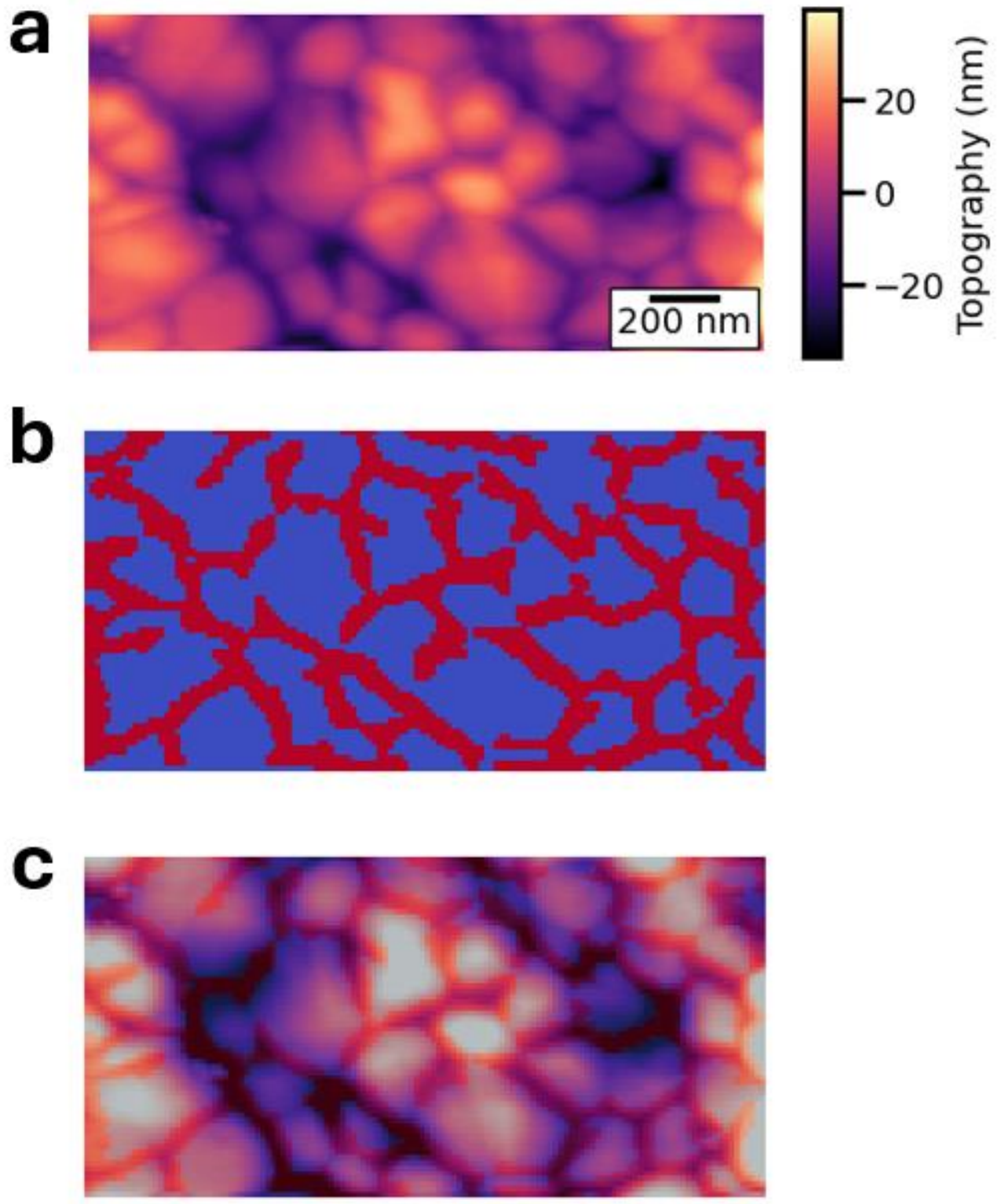

**(a)** Representative topography of illumination bias region of interest shown in Fig. 4 of the main text; **(b)** binary mask calculated from topography representing grain boundaries (red) and grain interiors (blue); and **(c)** binary mask overlaid on topography.

**Supplementary Fig 11.** Binary mask used to separate grain interiors and boundaries for second region of interest.

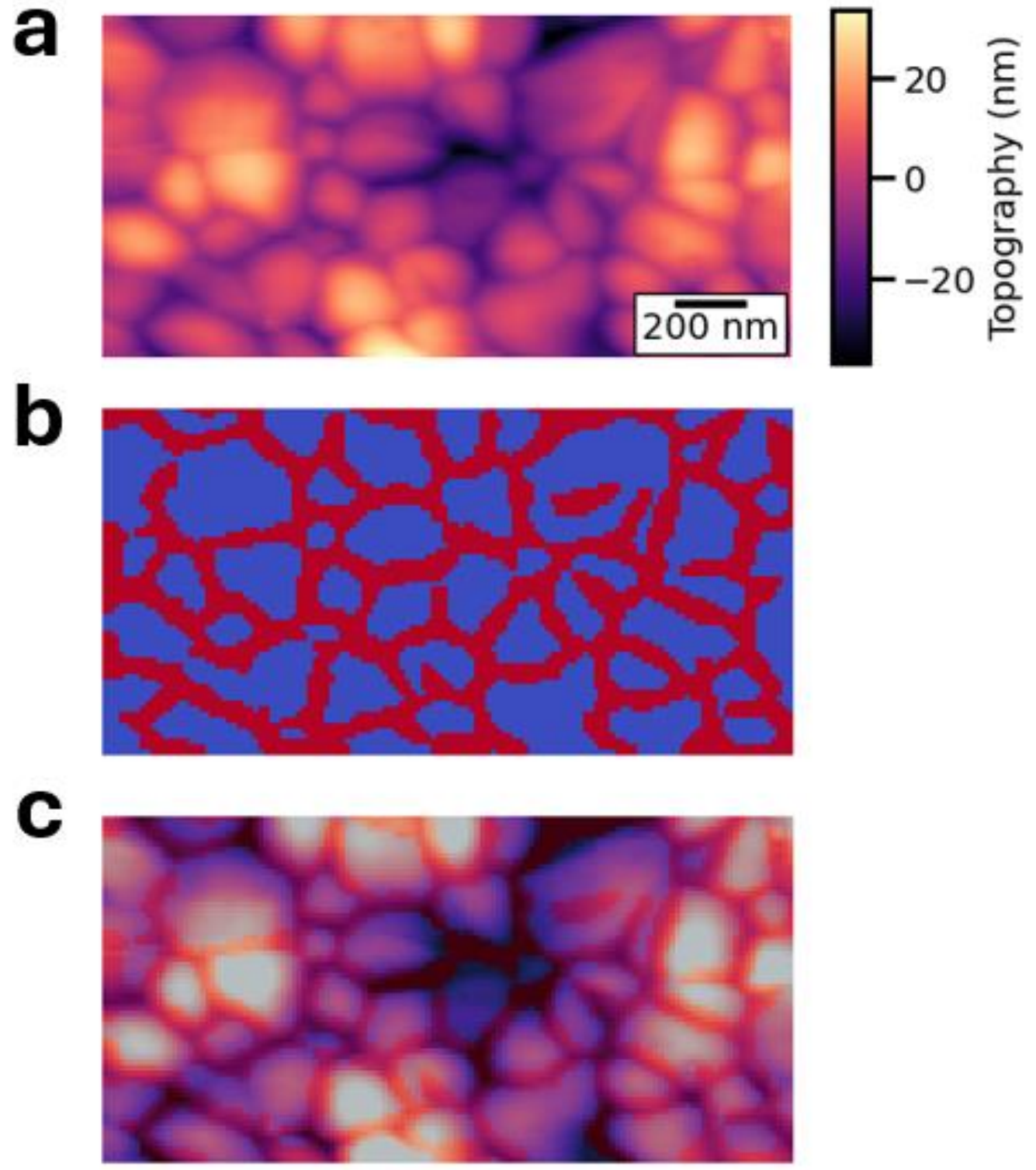

**(a)** Representative topography of illumination bias *second* region of interest (see Supplementary Fig. 12); **(b)** binary mask calculated from topography representing grain boundaries (red) and grain interiors (blue); and **(c)** binary mask overlaid on topography.

**Supplementary Fig 12.** Illumination bias data for second region of interest (same illumination conditions as main text Fig. 4).

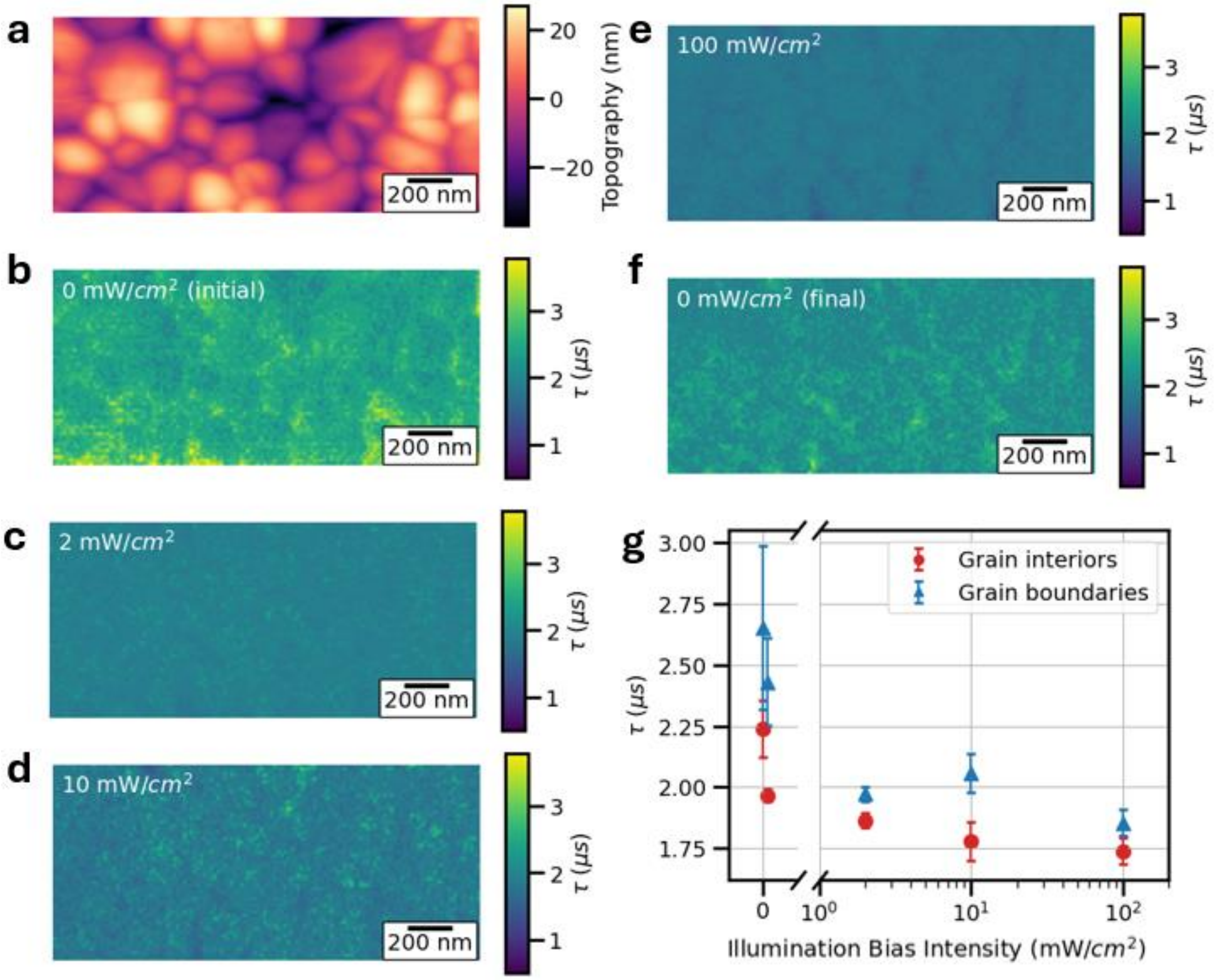

**(a)** Representative topography of second ROI, showing nanoscale morphology; **(b)-(e)** trEFM surface potential equilibration time images of the same ROI with increasing 660 nm bias illumination intensity ranging from 0 mW/cm2 to 100 mW/cm2, where a fast 405 nm laser at 110 mW/cm2 was used for trEFM excitation; **(f)** reproduced trEFM surface potential equilibration time image at 0 mW/cm2 showing return to unbiased time constants; **(g)** average grain interior and boundary trEFM surface potential equilibration times with respect to illumination bias intensity, error bars show standard deviation of masked pixel selection. Supplementary Fig. 11 shows binary mask used to extract grain boundary and interior data. Mean grain interior trEFM surface potential equilibration times decrease 22.1% at 100 mW/cm$^2$ with respect to unbiased image, and mean grain boundary trEFM time constants decrease 27.0%.

**Supplementary Fig 13.** Binary mask used to separate grain interiors and boundaries for third region of interest.

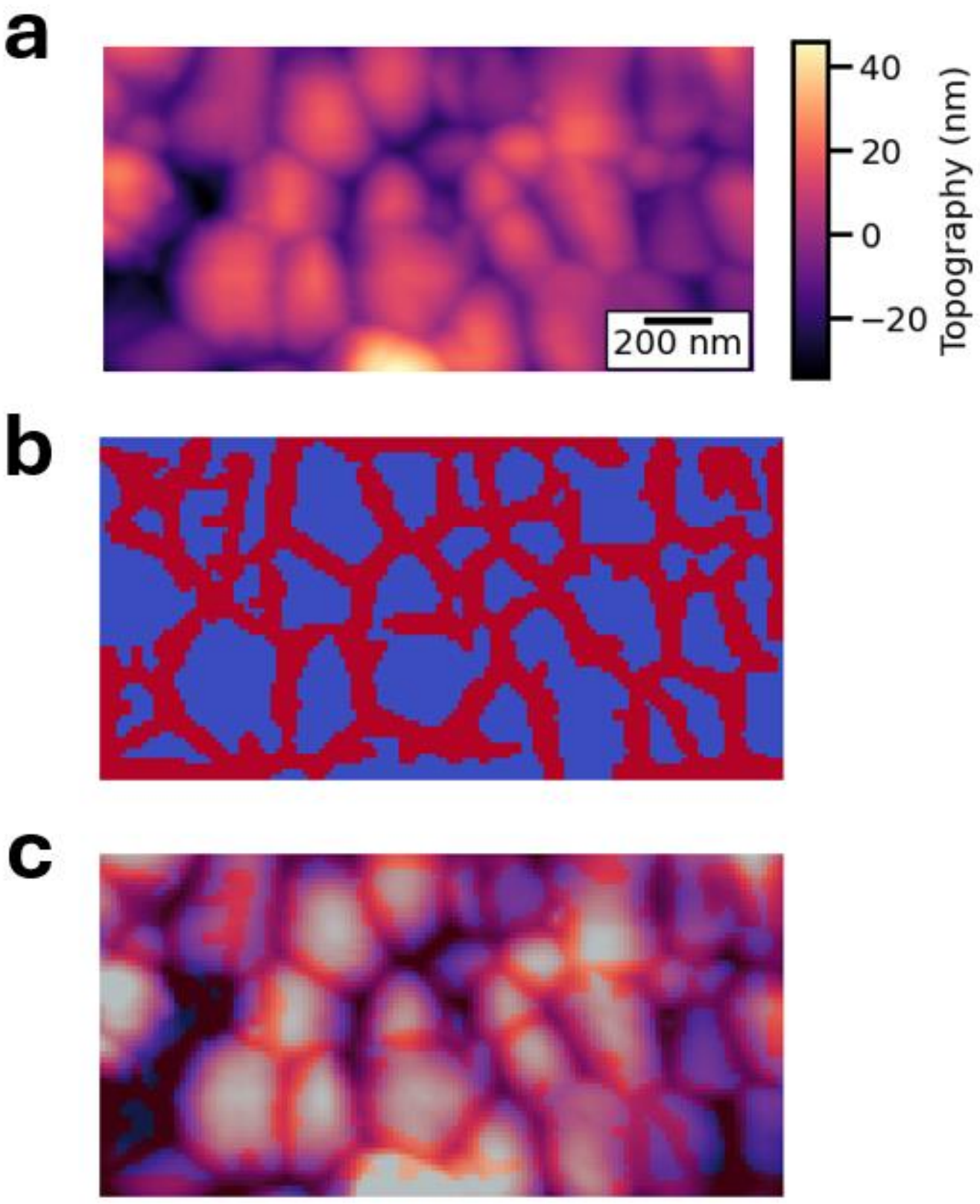

**(a)** Representative topography of illumination bias *third* region of interest where sample was biased with variable intensity 455 nm light and excited with the fast 405 nm laser (see Supplementary Fig. 14); **(b)** binary mask calculated from topography representing grain boundaries (red) and grain interiors (blue); and **(c)** binary mask overlaid on topography

**Supplementary Fig 14.** Illumination bias data for third region of interest.

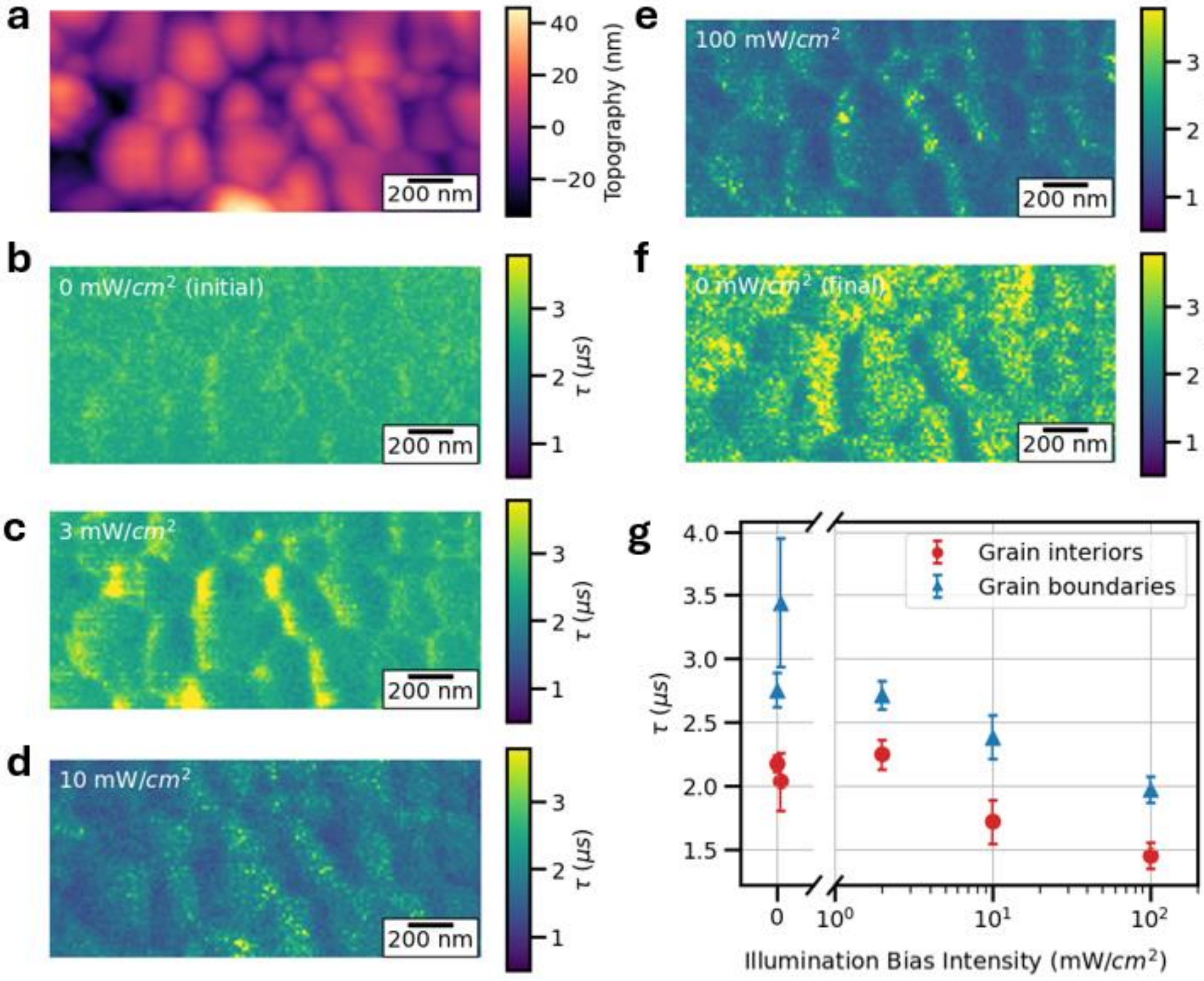

**(a)** Representative topography of third ROI, showing nanoscale morphology; **(b)-(e)** trEFM surface potential equilibration time images of the same ROI with increasing 455 nm bias illumination intensity ranging from 0 mW/cm2 to 100 mW/cm2, where a fast 405 nm laser at 110 mW/cm2 was used for trEFM excitation; **(f)** reproduced trEFM surface potential equilibration time image at 0 mW/cm2 showing return to unbiased surface potential equilibration times; **(g)** average grain interior and boundary trEFM surface potential equilibration times with respect to background illumination bias intensity, error bars show standard deviation of masked pixel selection. Supplementary Fig. 13 show the mask used for grain boundary and interior analysis. Mean grain interior trEFM surface potential equilibration times decrease 17.6% at 100 mW/$cm^2$ with respect to unbiased image, and mean grain boundary trEFM surface potential equilibration times decrease 31.2%.

**Supplementary Fig 15.** Excitation intensity dependent trPL lifetimes.

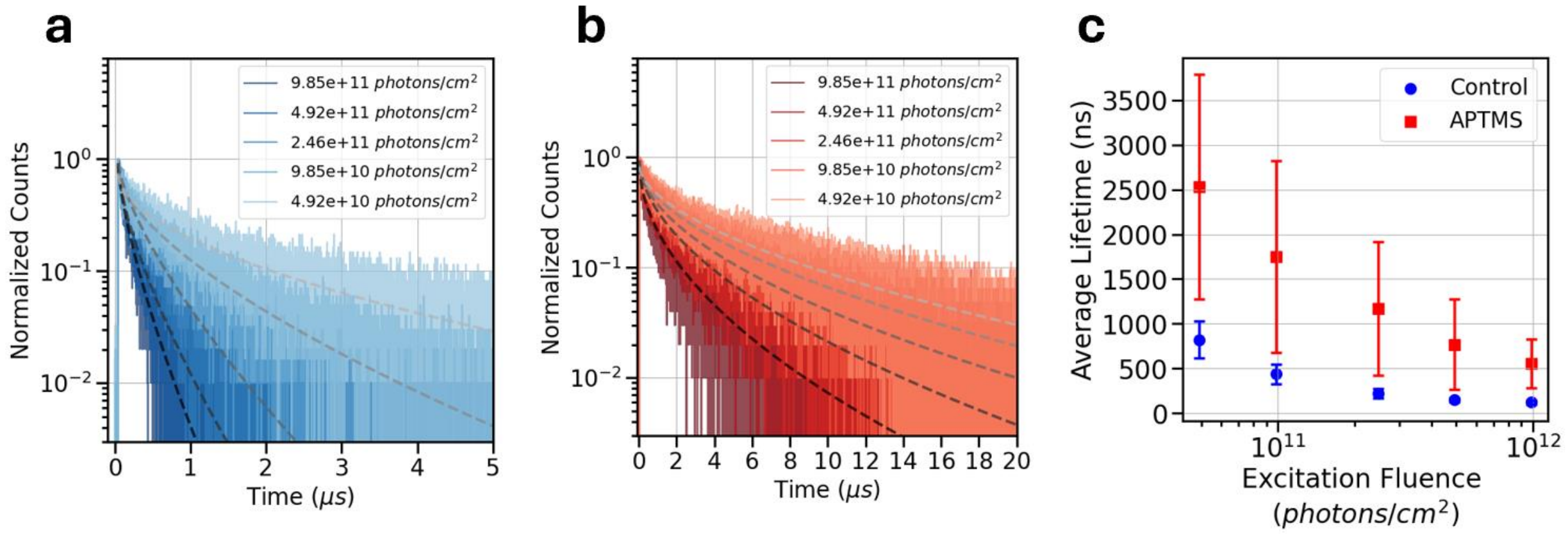

**(a)** Representative measured trPL decays for a $Cs_{0.22}FA_{0.78}Pb(I_{0.85}Br_{0.15})_3$ half-stacks before and **(b)** after APTMS passivation with varied excitation fluences. Stretched exponential fits are shown with the dashed lines. **(c)** Average trPL lifetime measured as a function of the excitation intensity of a 640 nm laser for a set of $Cs_{0.22}FA_{0.78}Pb(I_{0.85}Br_{0.15})_3$ half-stacks before and after passivation. The average value is calculated from measurements made on three separate samples and the error bars represent the standard deviation of those three samples. See Supplementary Table 4 for fitting parameters.

**Supplementary Fig 16.** Excitation intensity dependent trEFM measurements.

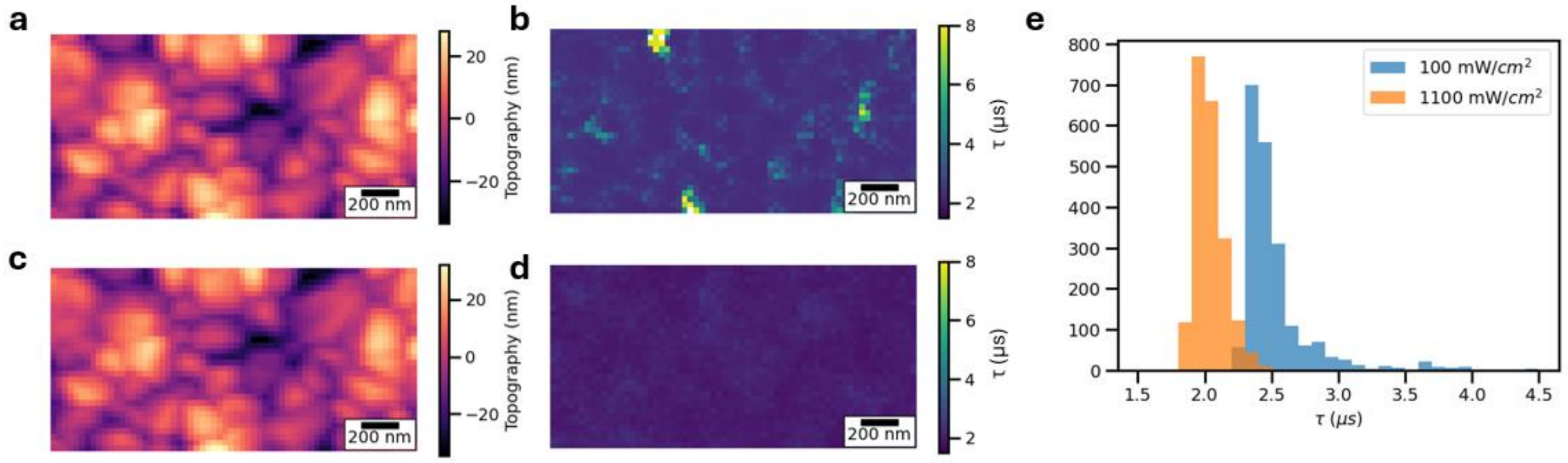

**(a)** Representative topography of a $Cs_{0.22}FA_{0.78}Pb(I_{0.85}Br_{0.15})_3$ half-stack showing nano-scale grains; **(b)** trEFM surface potential equilibration time map from 100 mW/cm$^2$ excitation with 405 nm laser showing slower dynamics at grain boundaries, as expected (see main text Fig. 2d); **(c)** topography of the same region of interest; **(d)** trEFM surface potential equilibration time map from image taken with 1100 mW/cm$^2$ 405 nm excitation, showing no observed contrast at grain boundaries and faster dynamics overall; and **(e)** histogram of these two trEFM surface potential equilibration time images validating our observation of faster (and lower contrast) surface potential equilibration dynamics captured with higher excitation intensity; at higher excitation intensity, higher order recombination dynamics have a higher contribution to overall electronic carrier recombination, meaning the sample reaches its equilibrium condition at earlier times.

**Supplementary Fig 17.** IonMonger excitation intensity simulations.

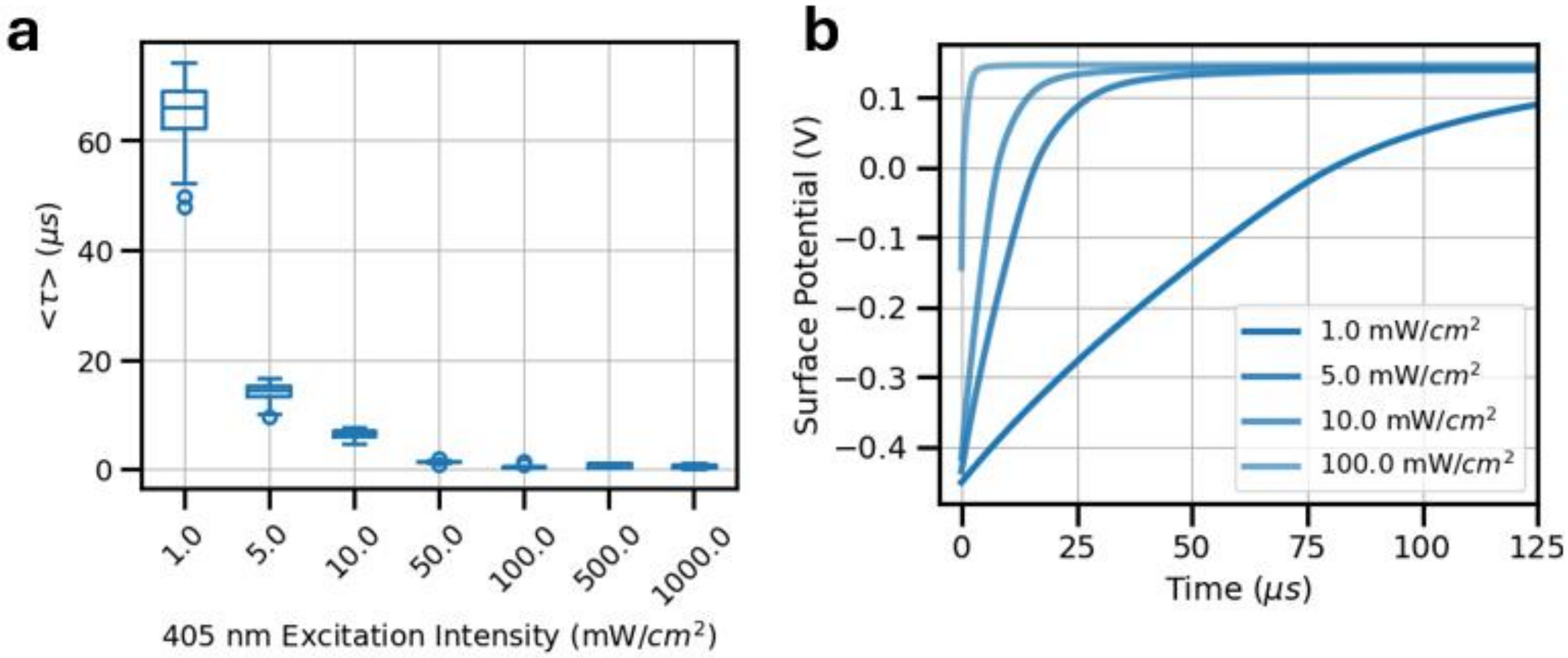

(a) Boxplot showing <τ> values calculated from tri-exponential fits of simulated surface potential dynamics plotted against excitation intensity with 405 nm excitation, showing faster dynamics when absorber is excited with higher intensity light; (b) shows example surface potential traces with varied excitation intensities, demonstrating faster dynamics with higher excitation intensity. For full simulation details, see Supplementary Table 3, where primary parameter swept was Illumination Intensity (scalar).

**Supplementary Fig 18.** Wavelength dependent trEFM measurements.

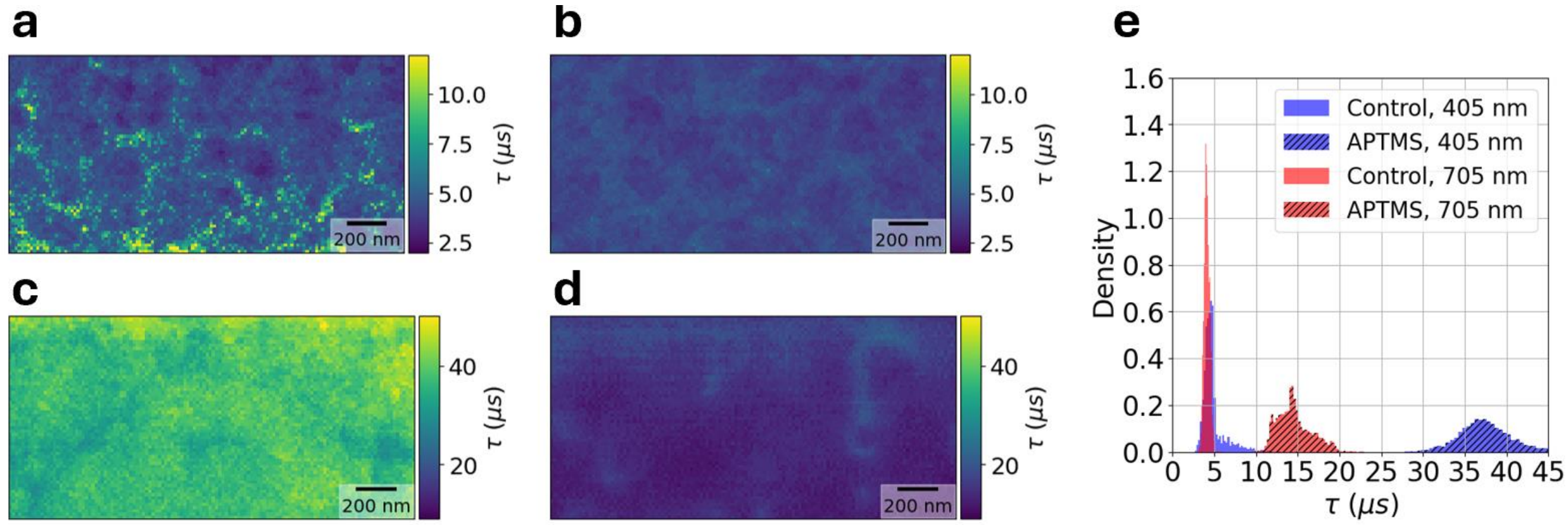

**(a)** Representative trEFM surface potential equilibration time images for $Cs_{0.22}FA_{0.78}Pb(I_{0.85}Br_{0.15})_3$ control half-stacks excited with 405 nm and **(b)** 705 nm fast lasers. **(c)** Representative trEFM surface potential equilibration time images for APTMS-treated half-stacks excited with 405nm and **(d)** 705 nm fast lasers. All images were collected with an incident intensity of ~150 mW/cm$^2$. Note that **(a)** and **(b)** share a color bar and **(c)** and **(d)** share a color bar to show the difference observed with different wavelengths. **(e)** Histogram representations of all images where 405 nm excitation, 705 nm excitation, and APTMS-treatment correspond to blue, red, and hashed appearance, respectively. Excitation with 705 nm light yields faster surface potential equilibration times compared to excitation with 405 nm light.

**Supplementary Fig 19.** IonMonger wavelength dependence simulations.

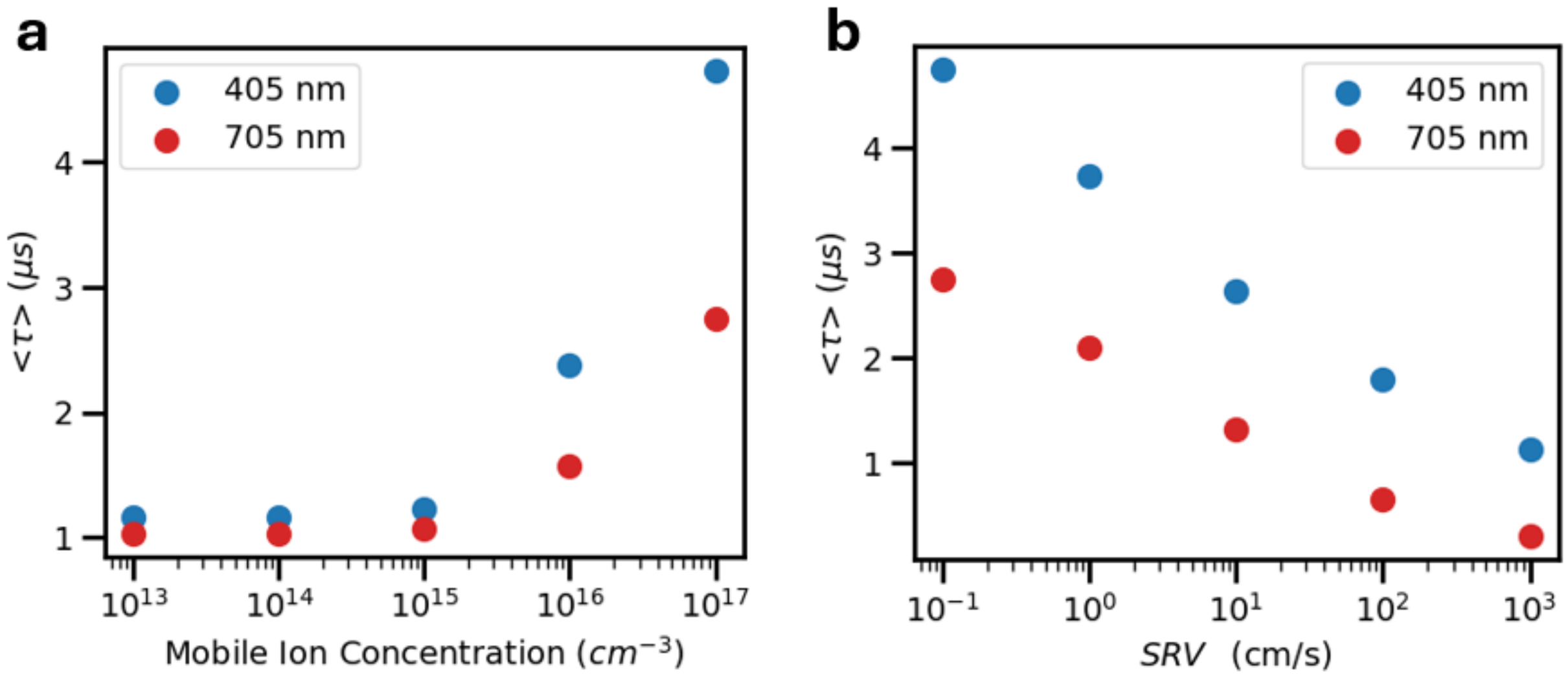

**(a)** <τ> values calculated from a tri-exponential fit to simulated surface potential evolution, plotted against variable mobile ion concentrations for 405 nm and 705 nm excitation, showing 705 nm excitation consistently results in faster surface equilibration dynamics; and **(b)** showing that at varied surface recombination velocities, 705 nm excitation still results in faster equilibration dynamics than 405 nm excitation. For full simulation details, see Supplementary Table 3, where the primary parameters swept were mobile ion concentration and surface recombination velocity, with all other parameters held constant.

**Supplementary Table 4.** Excitation fluence dependent stretched exponential fitting parameters for Glass/ITO/Me-4PACz/$Cs_{0.22}FA_{0.78}Pb(I_{0.85}Br_{0.15})_3$ half-stack samples before and after passivation with APTMS. Representative for Supplementary Fig 15a,b.

| Sample, Fluence (photons/cm$^2$) | $\tau_C$ (ns) | β-factor |
|---|---|---|
| **Control, 9.85e11** | 92 | 0.72 |
| **Control, 4.92e11** | 118 | 0.69 |
| **Control, 2.46e11** | 209 | 0.72 |
| **Control, 9.85e10** | 390 | 0.65 |
| **Control, 4.92e10** | 566 | 0.54 |
| **APTMS, 9.85e11** | 504 | 0.52 |
| **APTMS, 4.92e11** | 853 | 0.54 |
| **APTMS, 2.46e11** | 1261 | 0.54 |
| **APTMS, 9.85e10** | 1929 | 0.57 |
| **APTMS, 4.92e10** | 2483 | 0.57 |

**References.**